\documentclass[twocolumn]{aastex701}

\usepackage{CJK}
\usepackage{longtable}
\usepackage{threeparttable}
\usepackage{xspace}
\usepackage{amsmath}

\AtBeginDocument{
    \setcounter{topnumber}{1}      
    \setcounter{dbltopnumber}{1}   
    \setcounter{totalnumber}{1}    
    }

\usepackage[dvipsnames]{xcolor}

\newcommand{\Teff}{\ensuremath{T_{\mathrm{eff}}}\xspace}
\newcommand{\vsini}{\ensuremath{v{\mathrm{sin}i_*}}\xspace}
\newcommand{\veq}{\ensuremath{v_{\mathrm{eq}}}\xspace}
\newcommand{\Prot}{\ensuremath{P_{\mathrm{rot}}}\xspace}
\newcommand{\Rstar}{\ensuremath{R_*}\xspace}
\newcommand{\Rsun}{\ensuremath{R_\odot}\xspace}
\newcommand{\Mstar}{\ensuremath{M_*}\xspace}

\newcommand{\Mpl}{\ensuremath{M_{\mathrm{p}}}\xspace}
\newcommand{\msini}{\ensuremath{m{\mathrm{sin}i}}\xspace}
\newcommand{\Mjup}{\ensuremath{M_\mathrm{J}}\xspace}
\newcommand{\ecc}{\ensuremath{e_{\mathrm{orb}}}\xspace}
\newcommand{\aorb}{\ensuremath{a_{\mathrm{orb}}}\xspace}

\newcommand{\istar}{\ensuremath{i_*}\xspace}
\newcommand{\iorb}{\ensuremath{i_{\mathrm{orb}}}\xspace}
\newcommand{\Deltai}{\ensuremath{\Delta i}\xspace}
\renewcommand{\deg}{\ensuremath{^\circ}\xspace}

\providecommand{\NTotStars}{61\xspace}
\providecommand{\NTotComp}{69\xspace}

\providecommand{\NCirc}{17\xspace}
\providecommand{\NEcc}{52\xspace}

\providecommand{\NMisAll}{19\xspace}
\providecommand{\NMisCirc}{0\xspace}
\providecommand{\NMisEcc}{19\xspace}

\providecommand{\PerMisAll}{28$^{+6}_{-5}$\%\xspace}
\providecommand{\PerMisCirc}{0$^{+8}_{-0}$\%\xspace}
\providecommand{\PerMisEcc}{37$^{+7}_{-6}$\%\xspace}

\providecommand{\PerMisEccCJs}{41$\pm$8\xspace}

\providecommand{\AverageDeltai}{33.5$^\circ$}
\providecommand{\AverageDeltaiSigma}{12.0$^\circ$}

\providecommand{\NMultiHosts}{13\xspace}

\providecommand{\MinCompSep}{20\xspace}
\providecommand{\MaxCompSep}{5449\xspace}

\providecommand{\NMisMulti}{4\xspace}
\providecommand{\NMultiComp}{13\xspace}
\providecommand{\PerMisMulti}{31$^{+13}_{-11}$\%\xspace}
\providecommand{\NMisSingle}{15\xspace}
\providecommand{\NSingleComp}{56\xspace}
\providecommand{\PerMisSingle}{27$\pm$6\%\xspace}

\begin{document}


\title{Stellar Obliquities of Cold Jupiter and Brown Dwarf Systems I: Spin–Orbit Misalignment Grows with Orbital Eccentricity}

\author[0000-0003-2646-3727]{Lauren Biddle}
\altaffiliation{These authors contributed equally to this work}
\email[show]{lauren.biddle@princeton.edu}
\affiliation{Department of Astrophysical Sciences, Princeton University, 4 Ivy Lane, Princeton, NJ 08540, USA}

\author[0000-0002-2696-2406]{Jingwen Zhang(\begin{CJK*}{UTF8}{gbsn}张婧雯\end{CJK*})}
\altaffiliation{These authors contributed equally to this work}
\email[show]{jwzhang@UCSB.edu}
\affiliation{Department of Physics, University of California, Santa Barbara, CA 93106, USA}

\correspondingauthor{Lauren Biddle \& Jingwen Zhang}


\begin{abstract}

Stellar obliquity provides insights into the formation and dynamical evolution of planetary systems. Most obliquity measurements to date have focused on close-in transiting giant planets, whose orbits may have been altered by migration and tidal interactions. In contrast, the stellar obliquities of cold Jupiters, the most common class of giant planets, remain largely unexplored. Here, we measure the minimum stellar obliquity for \NTotComp cold Jupiters and brown dwarfs ($\rm M_p>0.3\ M_J$, $\rm a>1~au$), by comparing stellar inclinations derived from stellar rotation measurements with companion orbital inclinations constrained by radial velocities, relative astrometry and \textit{Hipparcos}--\textit{Gaia} absolute astrometry. We find that spin--orbit misalignments are common among cold Jupiters and brown dwarf companions, with \NMisAll/\NTotComp\ (\PerMisAll) exhibiting significant misalignments. Furthermore, the stellar obliquities depends strongly on orbital eccentricity, whereby \NMisCirc/\NCirc (\PerMisCirc) companions with \ecc$\leq$0.15 are misaligned, compared to \NMisEcc/\NEcc (\PerMisEcc) with more eccentric orbits. 
This pattern reveals a connection between orbital eccentricity and stellar obliquity: companions on nearly circular orbits tend to be aligned with their host stars, whereas those on eccentric orbits exhibit a broader range of obliquities. This finding suggests that post-formation dynamical interactions, such as planet--planet scattering or secular chaos, may play an important role in exciting both orbital eccentricities and spin--orbit misalignments.

\end{abstract}




\section{Introduction}\label{sec:intro}

Stellar obliquity, also known as spin-orbit angle, measures the angle between a star's spin vector and the orbital angular-momentum vector of its planets. It preserves a fossil record of the planetary system's formation and subsequent dynamical evolution. Modern helioseismology has measured the angle between the ecliptic plane and the solar equator with high precision as $7^{\circ}.155\pm{0.002}$ \citep{Beck2005}. The low solar obliquity is considered to inherited from the protoplanetary disk  and has been maintained throughout the history of the Solar System \citep{Kant1775, laplace1796, Hayashi1985, Lissauer1993}. 

To date, sky-projected stellar obliquities have been measured for hundreds of hot Jupiter systems ($<\sim$0.1 AU), primarily through the Rossiter--McLaughlin (RM) effect \citep{M1924,R1924}. This effect produces an anomalous radial velocity (RV) signal during planetary transits as the planet sequentially occults different regions of the rotating stellar disk. Previous studies have shown that a substantial fraction of hot Jupiters are strongly misaligned with their host stars, with the occurrence of misalignment showing a dependence on stellar effective temperature \citep{Winn2010,Schlaufman2010}. Hot Jupiters around cool stars ($T_{\rm eff}$~$<$~6250~K) are mostly aligned, potentially from efficient tidal realignment as a result of their thick convective envelopes \citep{Winn2010, Albrecht2012}. In contrast, hot Jupiters around hotter stars, where tidal dissipation is less efficient, exhibit a much broader range of stellar obliquities \citep{Winn2010, Schlaufman2010}. The stellar obliquities of close-in hot Jupiters are therefore imprinted by a mix of processes beginning with the primordial stellar obliquity inherited from the protoplanetary disk upon formation, then dynamical interactions with other planets or stars, the migration process, and later tidal realignment.  Disentangling the contributions of these processes remains challenging for close-in giant planets.

Stellar obliquity measurements have recently been extended to warm Jupiters at orbital separations of $\sim$0.1--1 au, where impact of tides becomes negligible. Rossiter--McLaughlin observations reveal that warm Jupiters preferentially orbit in alignment with their host stars, even around hot stars \citep{Rice2022b, Wang2024}, suggesting that warm Jupiters mostly form in aligned orientations and undergo limited subsequent dynamical evolution compared to hot Jupiters. However, independent constraints from stellar inclination measurements indicate that at least 24\% of warm Jupiter around cool stars are misaligned--- comparable to misalignment rate  of hot Jupiters around similar hosts \citep{Morgan2024, Biddle2026}.

Stellar obliquity measurements have also been extended to substellar companions \citep{Bowle2023} and protoplanetary disks \citep{Biddle2025} at wider separations of tens to hundreds of au, revealing that substantial spin--orbit misalignments are not uncommon in these systems. More recently, \citet{Biddle2026} compared line-of-sight stellar obliquities across populations of hot Jupiters, warm Jupiters, debris disks, and protoplanetary disks, finding that the obliquity distributions are largely similar. This similarity suggests that spin--orbit architectures may be established during formation, with limited subsequent dynamical evolution. However, these populations span substantially different orbital separations and evolutionary stages, from close-in planets around mature stars ($\sim$1--5~Gyr) to protoplanetary disks extending beyond 10~au around young stars ($\sim$2--5~Myr), complicating efforts to disentangle primordial misalignment from subsequent dynamical evolution.

Cold Jupiters at orbital separations of $\sim$1--10~au remain largely unexplored, despite representing the most common population of giant planets \citep{Fernandes2019,Fulton2021}. Unlike hot Jupiters, cold Jupiters are less likely to have undergone substantial inward migration and are largely unaffected by stellar tides. Their spin--orbit architectures  therefore preserve pristine information about their formation conditions and early dynamical evolution. In this work, we address this gap by placing stellar obliquity constraints on cold Jupiters and brown dwarf companions with orbital separations spanning $\sim$1--300~au. We infer stellar inclinations from rotation periods, stellar radii, and spectroscopic \vsini measurements, and determine companion orbital inclinations by combining radial velocities, absolute astrometry from Hipparcos and Gaia, and, where available, relative astrometry from direct imaging. This paper is organized as follows: Section \ref{sec:the_sample} describes the construction of the sample. Sections \ref{sec:stellar_inclinations}, \ref{sec:companion_orbits}, and \ref{sec:Deltai} outline the methods used to derive stellar and orbital inclinations and minimum obliquities, respectively. The results of this work are presented in Section \ref{sec:results}. We discuss the results in Section \ref{sec:Discussion}. A summary of this research is given in Section \ref{sec:summary}.

\section{The Sample} \label{sec:the_sample}

\subsection{Sample Construction}\label{sec:sample_construction}

\begin{figure}[h]
    \centering
    \includegraphics[width=\linewidth]{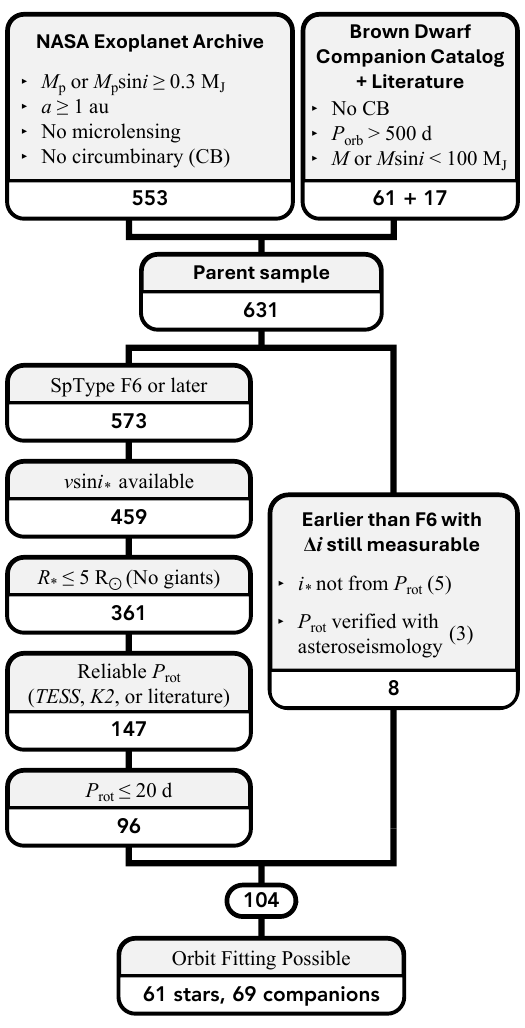}
    \caption{Flowchart illustrating the sample selection process.}
    \label{fig:fig_flowchart}
\end{figure}

Fig.~\ref{fig:fig_flowchart} presents the sample construction process. To construct the sample, we first selected stars from the NASA Exoplanet Archive \citep{Christiansen2025} that host giant companions (\Mpl$\geq$0.3~\Mjup) with orbit wide orbit semi-major axes (\aorb$\geq$1~au). We exclude the 146 companions discovered by microlensing. Their host stars lie at kpc distances and are too faint for stellar rotation measurements, and only the projected separation at the time of the event is known. We also remove companions flagged as circumbinary (20) to control for dynamical effects that can complicate the interpretation of the obliquity distribution of the sample. To this sample of 553 NEA cold Jupiter host stars, we add 61 non-circumbinary brown dwarf host stars from the Brown Dwarf Companion Catalog (BDCC; \citealt{Stevenson2023}), selecting systems with orbital periods $>$500~d and companion mass or \msini values that are $<$100~\Mjup. Finally, we supplement the compilation with 17 additional brown dwarf companions found in the literature, resulting in a parent sample of 631 stars.

Next, we apply a filter to the parent sample to reject stars for which \Prot (and thereby \istar) cannot be reliably determined from periodic brightness modulations in a light curve. Specifically, we remove stars with spectral types earlier than F6, because periodic brightness signatures present in the light curve are often caused by $\delta$~Scuti-like pulsations that are difficult to tell apart from signatures caused by rotation (e.g., \citealt{Sepulveda2024}). Following the removal, we are left with 573 stars with spectral types spanning F6 to M8.

\subsection{Stellar \vsini \& Radii}\label{sec:stellar_vsini}
For each star in the parent sample, we compile \vsini measurements from the literature and adopt the weighted mean (Table~\ref{tab:star_parameters}). Prior to computing the weighted mean, we first remove outliers that fall greater than one standard deviation from the unweighted mean of the compiled values. Out of the 621 total compiled \vsini values, 77 dropped \vsini values deviated from the mean by less than 2$\sigma$, and 18 dropped values deviated from the mean by more than 2$\sigma$. To reduce the effect of systematics from various approaches in data processing and uncertainty estimation stemming from different sources, we apply a 5\% lower limit to each \vsini measurement uncertainty, limiting the likelihood of one measurement unfairly weighting the mean. Then, if two or more compiled \vsini have no corresponding uncertainty, we compute the average and standard deviation of those values, apply the 5\% lower limit on the error, and treat it as one measurement. We enact a similar treatment when multiple \vsini measurements are compiled from a single source such that one \vsini measurement per literature source contributes to the calculation of the weighted mean. Some \vsini are reported as upper limits; if the only measurement found is an upper limit, it is adopted as reported. Otherwise, upper limits do not contribute to the weighted mean. If the resulting weighted mean is less than 2~km/s, we adopt a 2~km/s upper limit on \vsini. Stars with no \vsini available are removed, leaving 459 stars remaining in the sample. We provide the compilation of \vsini values for each system in Table \ref{tab:compiled_vsini}, and plot \vsini as a function of \Teff in Figure \ref{fig:Teff_vsini_Prot} in the Appendix.

Stellar radii are adopted from the TESS Input Catalog (TIC). For cases where the TIC does not report an uncertainty on the stellar radius, we assume 16\%, corresponding to the 95\% quantile of uncertainty distribution of \Rstar in the TIC. \Rstar from the TIC typically fall within 7\% of asteroseismic radius measurements of the same stars \citep{Huber2017}, so we add a 7\% error term in quadrature to the initial uncertainty to assure reliability in the adopted stellar radii. We remove stars without radius measurement reported in the TIC as well as giants with \Rstar$>$5~\Rsun, bringing the sample to 361 stars.

\subsection{Stellar Rotation Periods}\label{sec:stellar_rotation_period}
For each star, we conduct a uniform rotation period search within photometric light curves obtained with \textit{TESS} \citep{Ricker2015} and \textit{K2} \citep{Howell2014}, where available (see Appendix~\ref{app:light_curves}). With the cleaned light curves, we conduct a generalized Lomb--Scargle periodogram \citet{zechmeister2009} with a period-search grid spanning 0.2~d to 1/2 the full light curve baseline, with a sampling resolution of 0.01~d. For the majority of the resulting periodograms, we adopt the signal corresponding to the strongest peak as the rotation period and estimate its uncertainty as the half-maximum width of that peak. The adopted \Prot value is the weighted mean of the individual light curve measurements. We consider a rotation period to be reliable when the periodic signal is consistently the primary signature identified in the periodogram of all available light curves, and the amplitude of the rotation period is clearly distinguishable above the point-to-point scatter in the light curve. 76 stars meet this requirement. 
Figure~\ref{fig:example_light_curves} presents the TESS light curves and corresponding periodograms for AF~Lep and HD~29375. The complete set of light curves and periodograms for the remaining targets is available as a figure set in the online journal.

For the remaining stars, \Prot could not be measured with \textit{TESS} or \textit{K2} as a result of either no observations being available or because the rotation period could not be confidently identified. For these stars, we adopt \Prot for 71 stars from the literature, requiring that \Prot is derived from an photometric time series analysis, or alternatively, from a time series assessment of magnetically sensitive lines, which are an additional metric for tracing stellar rotation, effective even for faint or slowly rotating stars (e.g., \citealt{Vaughan1981, SuarezMascareno2015}).

The stellar sample, consisting primarily of G, K, and M stars are likely to exhibit slow rotation as a result of stellar spin-down over time \citep{Skumanich1972}. For cases where a star is slowly rotating (\Prot$\gtrsim$20~d) or its axis is nearly pole-on ($\sim$90\deg), \vsini becomes difficult to measure reliably because the amount of spectral line broadening from stellar rotation becomes comparable to, or falls below, broadening from macroturbulence \citep{Gray2005, Doyle2014} or the resolution sensitivity of the spectrograph. Although the typical sector length of \textit{TESS} light curves already serve to restrict many stars to relatively short rotation period, to ensure that the adopted \vsini measurements are reliable across the entire sample, we apply a cut in rotation period to exclude stars with \Prot$\>$20~d, bringing the sample size to 96.

The method used to determine \istar from \Prot, \vsini, and \Rstar relies on the assumption that \Prot traces the rotational velocity at the equator of the star. However, it is possible that spots could exist at non-equatorial latitudes (e.g., \citealt{Strassmeier2009, Berdyugina2005}), which could bias the resulting \istar toward higher inclinations. To account for possible differential rotation of star spots at high latitudes, we inflate the uncertainties on \Prot with an error term ($\sigma_{\rm shear}$), which is added quadrature to the nominal \Prot uncertainty (e.g., \citealt{Bowler2023, Biddle2025}). We apply this uncertainty inflation method to all \Prot uncertainties derived in this work and adopted from the literature.

\subsection{Additional Systems}
To the filtered sample, we add eight additional systems with spectral types earlier than F6 for which \Deltai can still be measured. For five of these systems, the stellar inclination has already been determined using an alternative method that does not rely on rotation period (see Section \ref{sec:stellar_inclinations}). For two systems, the stellar rotation period has been distinguished from pulsations with asteroseismology (51~Eri; \citealt{Sepulveda2022}, HR~8799; \citealt{Sepulveda2023}). An in-depth assessment of the photometric time series of the final star (HD~206893; \citealt{Delorme2017}) shows a slightly progressing phase in the periodic signature, indicative of rotation. If the origin of the period is actually caused by pulsations, the effective temperature (\Teff=6477~K) is comparable to $\gamma$-Dor like variables, for which evidence suggests that the stellar oscillations are at or close to the rotational period \citep{Balona2011}. The addition of these stars increases the sample size to 104.

\subsection{Assessment of Orbital Fitting Viability}
For systems with available stellar inclination measurements from the previous steps, we further select companions for which the existing observations are sufficient to constrain their three-dimensional (3D) orbits. The orbital fitting methodology is described in Section~\ref{sec:orbit_fit}. We compile all publicly available radial velocity (RV) measurements and relative astrometry, as summarized in Tables~\ref{tab:rel_astr} and \ref{tab:rv_data}.

We exclude systems primarily for a few reasons. First, we exclude directly imaged planetary and brown dwarf companions at projected separations beyond 200 au, for which the available observations typically cover only a small fraction of the orbit and are insufficient to constrain the 3-D orbital orbits.  Second, we exclude RV-detected giant planets with orbital periods longer than 30 years whose orbital parameters remain poorly constrained, particularly those for which the RV observations reveal only a long-term trend or curvature rather than a well-characterized Keplerian orbit. Additionally, we exclude systems in which the companion is itself an unresolved binary rather than a single object.

\begin{figure*}[tp]
    \centering
    \includegraphics[width=\linewidth]{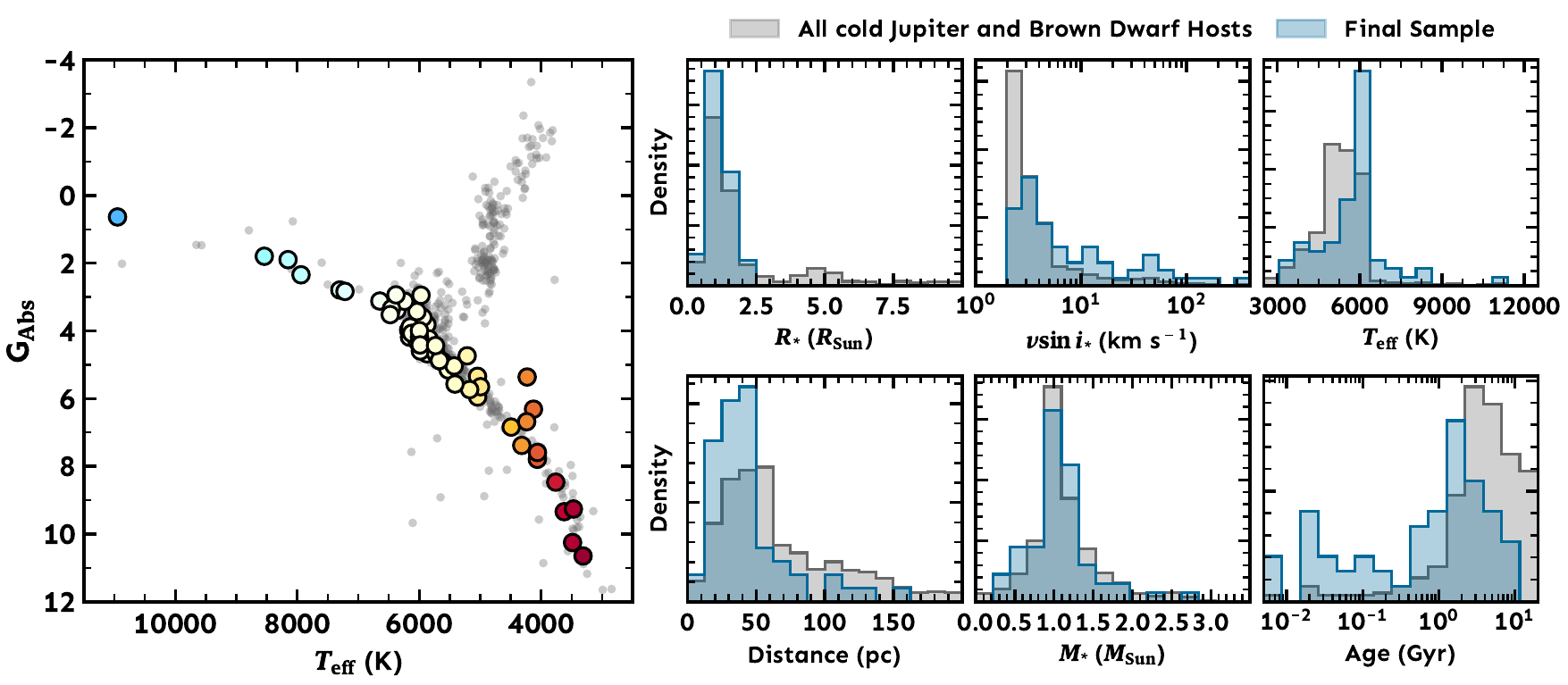}
    \caption{Overview of the stellar properties of our sample. \textit{Left}: Hertzsprung--Russell (HR) diagram ($T_{\rm eff}$--G$_{\rm Abs}$) of the host stars. Colored circles are host stars in the final sample. Gray points represent the parent sample of all confirmed cold Jupiter and brown dwarf host stars (see Section~\ref{sec:sample_construction}). \textit{Right Histograms}: Distributions of the stellar properties of the parent sample (gray) and the final sample (blue). The top row displays the distribution of $R_\star$, $v\sin i_\star$, and $T_{\rm eff}$. The bottom row displays stellar distance, mass \Mstar, and age.}
    \label{fig:HR_Diagram}
\end{figure*}

In summary, Our sample consists of stars that host at least one sub-stellar giant companion with well-characterized orbital inclinations, stellar radii (\Rstar), projected rotation velocities (\vsini), and rotation periods (\Prot). Our final sample consists of \NTotComp planets in \NTotStars planetary systems (see Figure \ref{fig:HR_Diagram}).



\section{Stellar Inclination \istar}\label{sec:stellar_inclinations}

To determine the stellar inclination, we combine \Rstar, \vsini, and \Prot to determine \istar. With this method, \Prot and \Rstar determine the equatorial rotational velocity (\veq, where \veq$=$2$\pi$\Rstar/\Prot), whereby the ratio \vsini/\veq provides an estimate of $\sin i_*$ and therefore the inclination of the stellar spin axis relative to the line of sight. We use the analytical expression for the posterior probability distribution of \istar presented in \citet{Bowler2023}, which follows the probabilistic framework of \citet{Masuda2020}, accounting for the correlation between \veq and \vsini. The analytic expression assumes an isotropic prior for \istar and uniform priors for \Prot, \Rstar, and \vsini:
\begin{equation}
P(i_* \mid P_{\rm rot}, R_*, v\sin i_*) \propto \sin i_* \times
\frac{
  e^{-\frac{\left(v\sin i_* - \frac{2\pi R_*}{P_{\rm rot}}\sin i_*\right)^2}
           {2\left(\sigma_{v\sin i_*}^2 + \sigma_{v_{\rm eq}}^2 \sin^2 i_*\right)}}
}{
  \sqrt{\sigma_{v\sin i_*}^2 + \sigma_{v_{\rm eq}}^2 \sin^2 i_*}
},
\end{equation}
\noindent where
\begin{equation}
\sigma_{v_{\rm eq}} = \frac{2\pi R_*}{P_{\rm rot}}
\sqrt{\left(\frac{\sigma_{R_*}}{R_*}\right)^2 + \left(\frac{\sigma_{P_{\rm rot}}}{P_{\rm rot}}\right)^2},
\end{equation}
and $\sigma_{\rm P_{rot}}$, $\sigma_{\Rstar}$, $\sigma_{\vsini}$ are the uncertainties of \Prot, \Rstar, and \vsini, respectively. The \vsini measurements for three stars (HD~147513, HD~171238, HIP~70849) are upper limits. For the upper limit case, we determine the posterior distribution of \istar with the analytical expression:
\begin{equation}
\begin{split}
P(i_* \mid P_{\rm rot}, R_*, v\sin i_*) \propto \sin i_* \times
\Bigg( & \mathrm{erf}\!\left(\frac{l - \frac{2\pi R_*}{P_{\rm rot}}\sin i_*}{\sqrt{2}\,\sigma_{v_{\rm eq}}\sin i_*}\right) \\
& + \mathrm{erf}\!\left(\frac{\sqrt{2}\,\pi R_*}{\sigma_{v_{\rm eq}} P_{\rm rot}}\right) \Bigg),
\end{split}
\end{equation}
where $l$ is the \vsini upper limit. 

For five stars, we adopt the stellar inclination from literature sources where \istar is measured using asteroseismology ($\beta$~Pictoris; \citealt{Zwintz2019}, HIP~65426; \citealt{Sepulveda2024}, 16~Cyg~B; \citealt{Davies2015}) or from stellar oblateness modeling of observed interferometric visibilities \citep{Jones2015} ($\kappa$~And; \citealt{Jones2016}, HIP~99770; \citealt{Balmer2026}).

\section{Planet/ Brown Dwarf Companion Orbits}\label{sec:companion_orbits}

We used the open-source package \textit{Octofitter} \citep{Thompson2023} to perform three-dimensional orbit fits. \textit{Octofitter} simultaneously models absolute astrometric measurements from \textit{Gaia} DR3, \textit{Gaia} DR2, and \textit{Hipparcos}, together with long-baseline radial velocities (RVs), relative astrometry from direct imaging, or both when available.

For most systems, we adopted a single-companion model. For each companion, we parameterized the orbit using the orbital period $P$, eccentricity $e$, argument of periastron $\omega$, cosine of the inclination $\cos i$, longitude of the ascending node $\Omega$, and orbital phase at a reference epoch $\theta$. We also fit for the host-star mass $M_*$ and companion mass $M_p$. Additional nuisance parameters included RV and astrometric noise terms, instrument-specific RV zero points, and astrometric epoch-selection parameters.  We assumed Gaussian priors on the host-star mass $M_{*}$ based on values reported in the literature. For the orbital parameters, we adopted a uniform prior on eccentricity between 0 and 0.99, and a uniform prior in $\cos i$ for the inclination. The priors on the argument of periastron $\omega$ and longitude of the ascending node $\Omega$ were uniform between 0 and $2\pi$. For the companion mass, we adopted a broad log-uniform prior spanning 0.1--100~$M_{\rm J}$.

\begin{figure*}[tp]
    \centering
    \includegraphics[width= 0.7\linewidth]{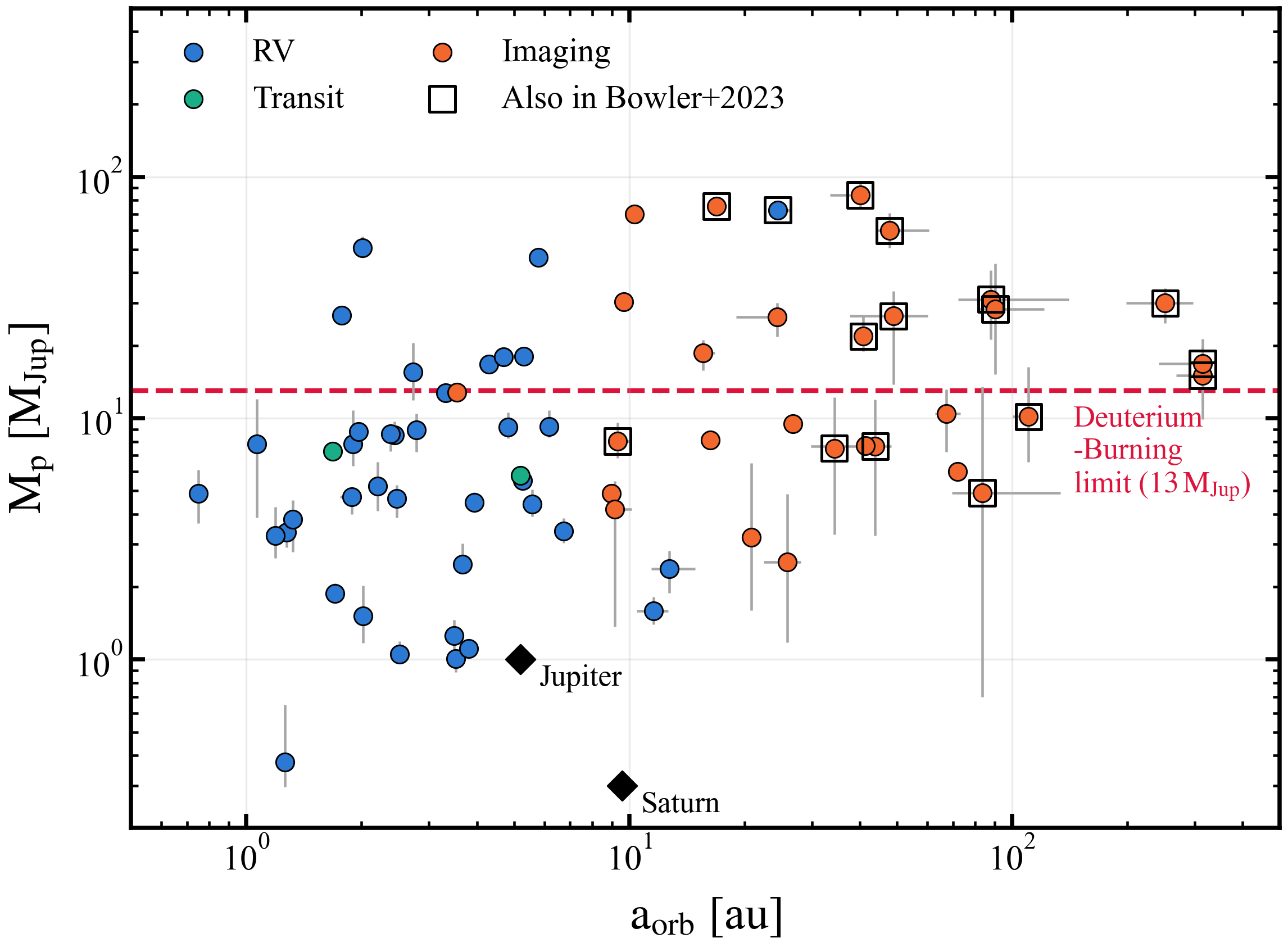}
    \caption{True masses and semi-major axes of the cold-Jupiter and brown-dwarf companions in our sample, taken from our 3D orbit fits. Each circle is colored by discovery method: radial velocities (orange), transits (green), and direct imaging (blue). Gray error bars show the $1\sigma$ uncertainties. Gray squares mark companions that also appear in \citet{Bowler2023}, who measured minimum stellar obliquities for a sample of directly imaged substellar companions. The dashed line shows the deuterium-burning limit ($\approx 13\,M_{\rm Jup}$), which we use to separate giant planets from brown dwarfs. Jupiter and Saturn (black diamonds) are shown for reference. }
    \label{fig:fig_planet_properties}
\end{figure*}

The total likelihood is constructed from the product of the RV, relative astrometry from imaging, absolute astrometry from \textit{Hipparcos} and \textit{Gaia}, and \textit{Gaia} astrometric excess noise (unbiased estimator of variance, or UEVA): 

\begin{equation}
ln\mathcal{L}
=
ln\mathcal{L}_{\mathrm{RV}}\,+
ln\mathcal{L}_{\rm{rel.\ ast.}}\,+
ln\mathcal{L}_{\mathrm{abs. \ ast. }}\,+
ln\mathcal{L}_{\mathrm{UEVA}}\,
\end{equation}

Absolute astrometric constraints from \textit{Hipparcos}, \textit{Gaia} DR2, and \textit{Gaia} DR3, together with \textit{Gaia} astrometric excess noise, were used for every system. For the absolute astrometry, we modeled a joint set of calibrated proper-motion observables: the \textit{Hipparcos} proper motion $\mu_{\mathrm H}$; the \textit{Hipparcos}--\textit{Gaia} long-baseline scaled position difference $\mu_{\mathrm{HG}}$; the \textit{Gaia} DR2 and DR3 proper motions, $\mu_{\mathrm{DR2}}$ and $\mu_{\mathrm{DR3}}$; and the \textit{Gaia} DR3--DR2 scaled position-difference term $\mu_{\mathrm{DR32}}$. The \textit{Hipparcos}-based quantities, $\mu_{\mathrm H}$ and $\mu_{\mathrm{HG}}$, were adopted from the DR3-calibrated HGCA solution. The \textit{Gaia} DR2 quantities were separately cross-calibrated onto the DR3 reference frame using a global magnitude-dependent rotation, local HEALPix offsets, and uncertainty inflation. We also included the DR2--DR3 cross-covariance to account for shared \textit{Gaia} transits between DR2 and DR3 (see \citet{Thompson2026} for details).

For planets detected through radial velocities, we adopted archival RV measurements when available. For directly imaged planets, we adopted archival relative astrometry from imaging observations. In systems where both RVs and relative astrometry were available, both data sets were included. Tables~\ref{tab:rel_astr} and \ref{tab:rv_data} provide a detailed summary of the RV and relative astrometry data, including the number of measurements and the observational baseline for each system.   We sampled the posterior using the stabilized variational nonreversible parallel tempering algorithm implemented in \textit{Pigeons.jl} \cite{pigeons} for $2^{12}$ scans. Figure \ref{fig:Example_Orbit_Fits} shows two representative examples of our orbital fits, combining absolute astrometry from \textit{Hipparcos}, \textit{Gaia} DR2, and \textit{Gaia} DR3 with either radial velocities or relative astrometry. Figure~\ref{fig:Example_Corner_Plots} shows the joint posterior distributions of the orbital parameters and companion masses for these two systems. The complete sets of orbital fits and corner plots for the remaining targets are available as figure sets in the online journal. Table ~\ref{tab:orbit_fit} presents the orbital fitting results for all the systems.

Figure~\ref{fig:fig_planet_properties} shows the companion masses and semi-major axes derived from our three-dimensional orbital fits. Our sample significantly expands previous stellar obliquity studies, particularly by probing cold Jupiters at separations of $\sim$1--10\,au, a regime largely unexplored by previous measurements. Our sample increases the number of companions beyond 1\,au with obliquity constraints from previous studies \citep{Bowle2023} by approximately a factor of three.

\section{Minimum Stellar Obliquity \Deltai}\label{sec:Deltai}

Stellar obliquity, also commonly referred to as spin-orbit angle, is the angle between the stellar spin axis and the normal vector of the orbital plane of planets. The true stellar obliquity angle $\psi$ is given by \citep{FW2009}:
\begin{equation}
\cos\Psi = \cos i_\star \cos i_{\rm orb} + \sin i_\star \sin i_{\rm orb} \cos{\lambda},
\end{equation}
where $i_\star$ and $i_{\rm orb}$ are the inclinations of the stellar spin axis and the companion's orbit, and $\lambda$ is the sky-projected stellar obliquity (see Fig.~\ref{fig:obliquity_geometry}). However, the position angle of the stellar spin axis on the sky, and hence $\lambda$, is generally unknown for non-transiting planets, so $\psi$ cannot be determined directly. Following \citet{Bowler2023}, we quantify spin--orbit alignment by determining the absolute difference between $i_*$ and $i_{\rm orb}$:
\begin{equation}
\Delta i = |i_\star - i_{\rm orb}|.
\end{equation}
This observable measures the relative tilt between stellar equatorial plane and companion's orbital plane along the line of sight, and provides a minimum value of the true stellar obliquity $\Psi$.

\begin{figure*}[tp]
    \centering
    \includegraphics[width=0.8\linewidth]{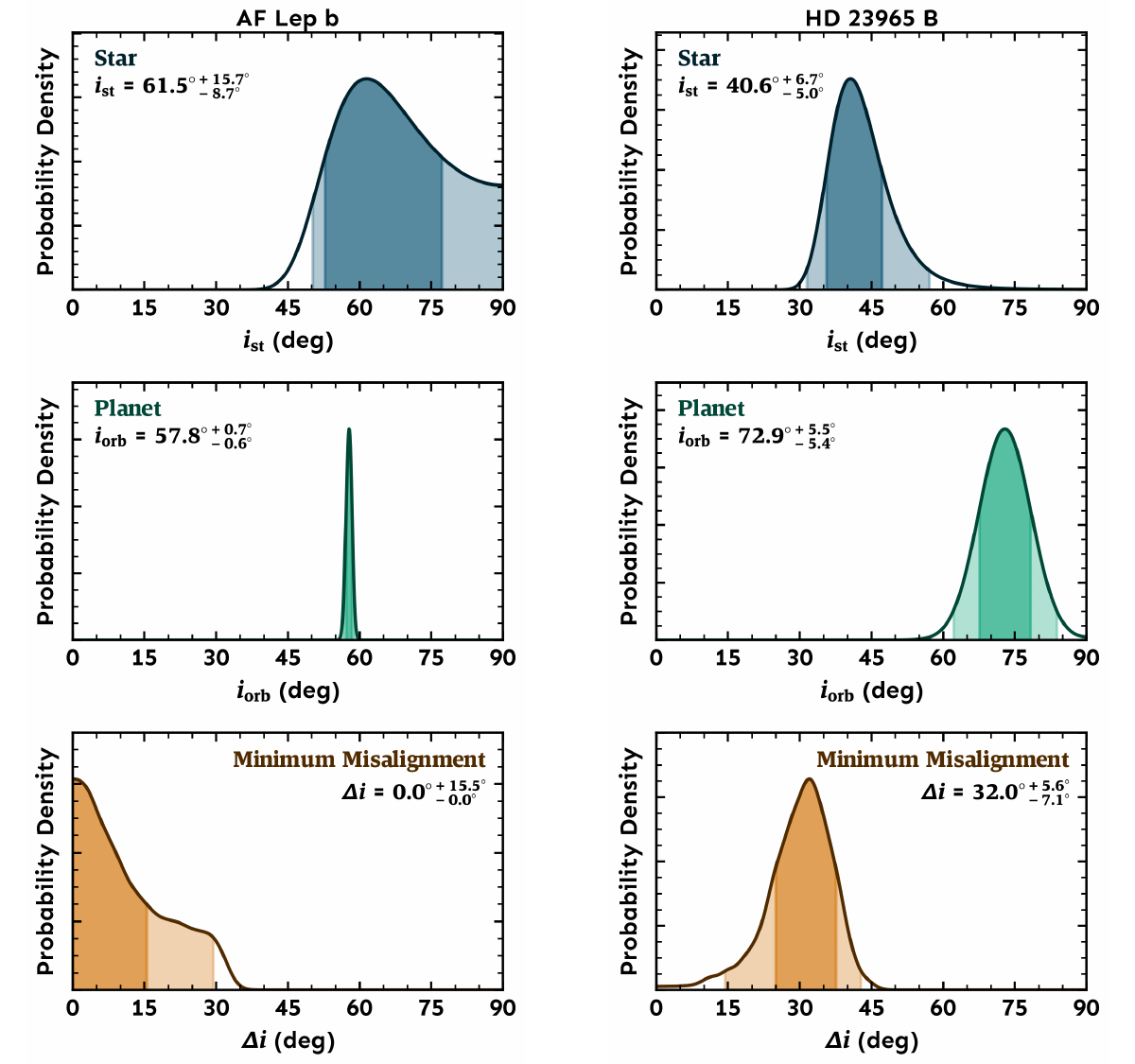}
    \caption{Posterior probability distributions of \istar (\textit{top}), \iorb (\textit{middle}), and \Deltai (\textit{bottom}) for AF~Lep~b and HD~23965~B. AF~Lep~b is an example where \Deltai is consistent with spin-orbit alignment, and HD~23965~B is an example of a misaligned system. The dark and light shaded regions indicate the $1\sigma$ and $2\sigma$ intervals around the maximum a posteriori (MAP) estimate, respectively. The complete figure set for all systems (69 images) is included in the submission materials and will be made publicly available upon publication of the paper.}
    \label{fig:example_posteriors}
\end{figure*}


Because \Deltai alone cannot distinguish prograde from retrograde configurations, we restrict both \istar and \iorb to the range $[0^\circ, 90^\circ]$, limiting the resulting obliquity estimate to the prograde solution. For some companions, the posterior distribution of \iorb is bimodal due to the prograde--retrograde degeneracy (see Figure~\ref{fig:fig_sideview}), with the two modes corresponding to the same orbital plane but opposite directions of orbital motion. Although these posteriors are generally approximately symmetric about $90^\circ$, we fold the $\iorb > 90^\circ$ solutions into the $\iorb < 90^\circ$ range using the transformation $\iorb \rightarrow 180^\circ - \iorb$. A detailed discussion of the prograde--retrograde degeneracy is provided in Appendix~\ref{sec:geometry}.

Figure~\ref{fig:example_posteriors} shows the posterior distributions of \istar, \iorb, and \Deltai for two representative companions, AF~Lep~b and HD~23965~B. To derive the posterior distribution of \Deltai for each star--companion pair, we randomly 15,000  samples from the \istar and \iorb posterior distributions and compute the absolute difference between the paired sample. We represent the \Deltai distribution with a boundary-corrected kernel density estimate (KDE) of the results. The bandwidth ($\kappa$) used to construct each \Deltai KDE is equal to the 16th--84th percentile distance of the set of \Deltai values, ranging from 0.13\deg ($\beta$~Pic~b) to 4.37\deg (HD~215497~c) across the sample. The adopted \Deltai value is the \textit{maximum a posteriori} (MAP) of the distribution and take the 68\% highest density interval of the distribution as the corresponding uncertainty. We list the adopted values for each companion in Table \ref{tab:inclinations_delta_i}.  The posterior distributions of \istar, \iorb, and \Deltai for the remaining systems will be made publicly available upon publication.

\begin{figure*}[tp]
    \centering
    \includegraphics[width=0.95\linewidth]{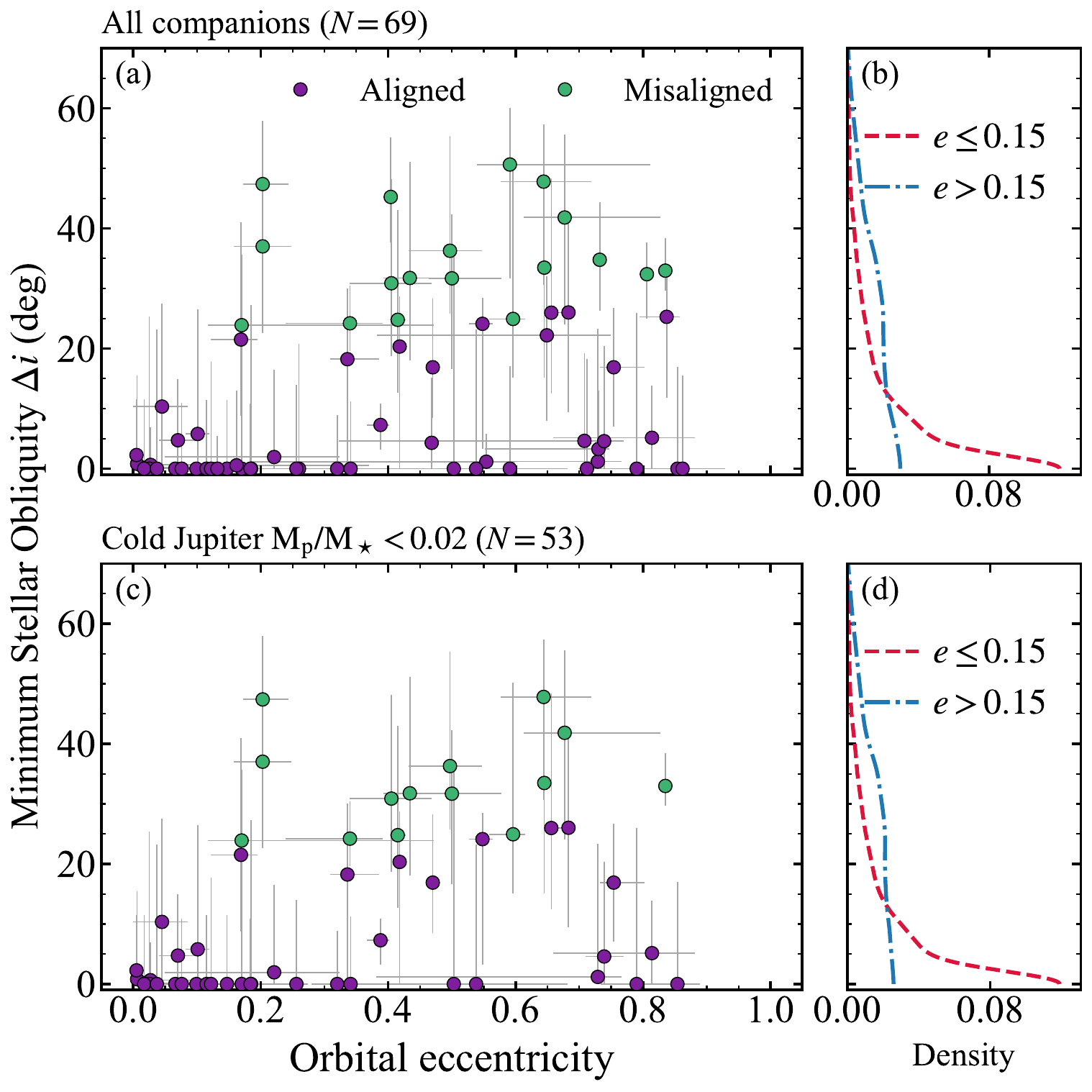}
    \caption{\textbf{Minimum stellar obliquity (\Deltai) as a function of the companion orbital eccentricity.} (a) Maximum a posteriori (MAP) estimates of $\Delta i=|i_\star-i_{\rm orb}|$, with $1\sigma$ measurement uncertainties for full sample. Green circles identify companions whose $95.4\%$ MAP-containing credible interval for $\Delta i$ excludes $0^\circ$; purple circles indicate those whose interval includes $0^\circ$. (b)  KDE distribution of full sample (black), companions with eccentricity below 0.15 (red) and those with eccentricity above 0.15 (blue).  (c)$\&$(d) are the same as (a)$\&$(b), but for cold Jupiter sample with mass ratio $\rm M_p/M_*$ below 0.02.}
    \label{fig:fig_delta_i_ecc}
\end{figure*}

\begin{figure}[tp]
    \centering
    \includegraphics[width=\linewidth]{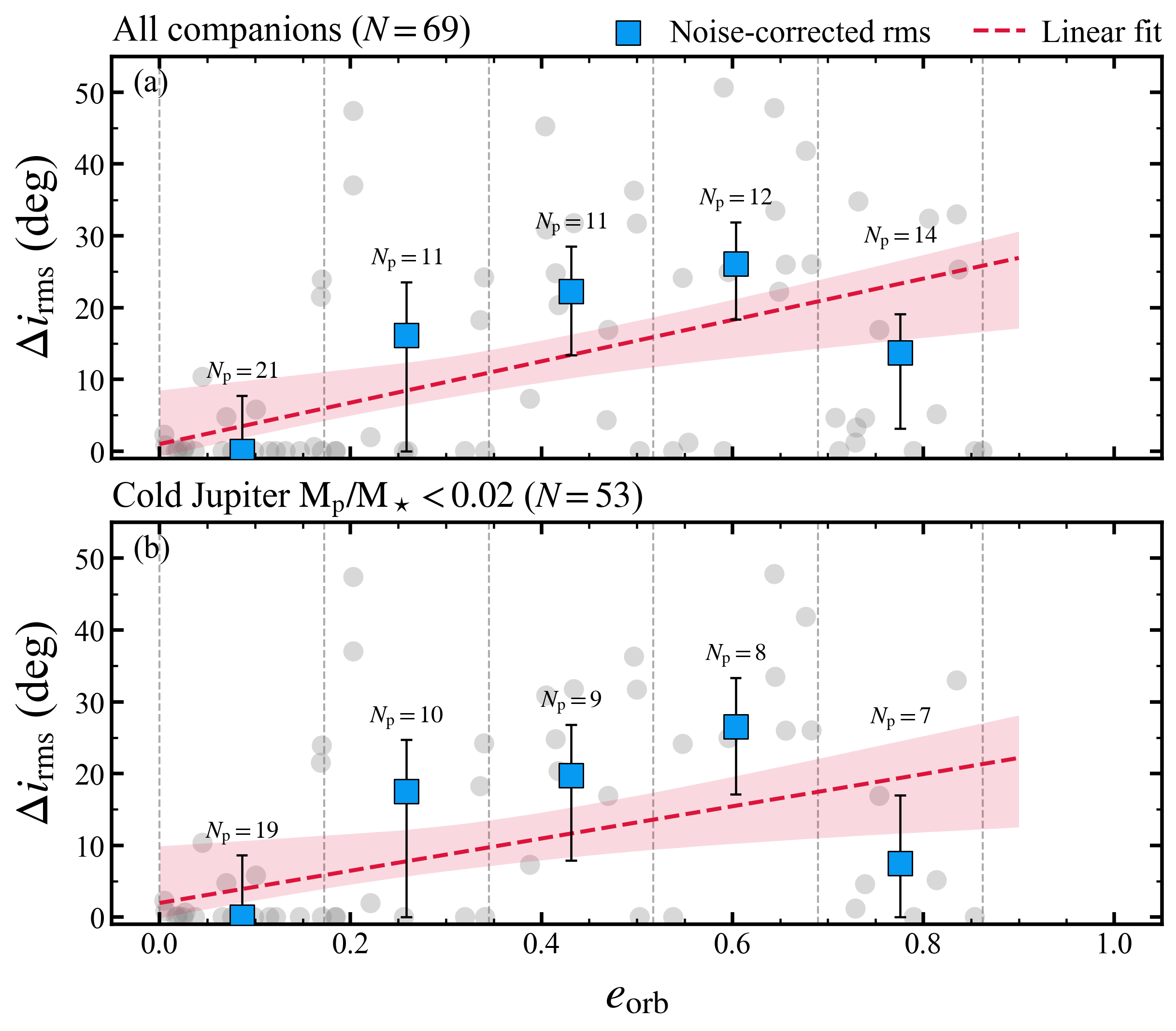}
    \caption{ Eccentricity dependence of the noise-corrected RMS inclination difference,
$\Delta i_{\rm rms}$, for (a) all 69 companions and (b) the 53 cold Jupiters
with companion-to-host mass ratio $M_{\rm p}/M_\star < 0.02$.
Gray circles show individual posterior MAP estimates of $\Delta i$ same in Fig.~\ref{fig:fig_delta_i_ecc}.
Blue squares show the noise-corrected RMS values in five equal-width
eccentricity bins; vertical dashed lines mark the bin boundaries, and
$N_{\rm p}$ denotes the number of companions in each bin.
Error bars span the 16th--84th percentiles of 20,000 bootstrap realizations
incorporating companion resampling and individual $\Delta i$ posterior
uncertainties. The red dashed line shows the fitted relation $\Delta i_{\rm rms}=k e_{\rm orb}$. with the shaded region representing its central $1\sigma$ posterior credible interval.}\label{fig:fig_delta_i_ecc2}
\end{figure}

\begin{figure}[tp]
    \centering
    \includegraphics[width=0.85\linewidth]{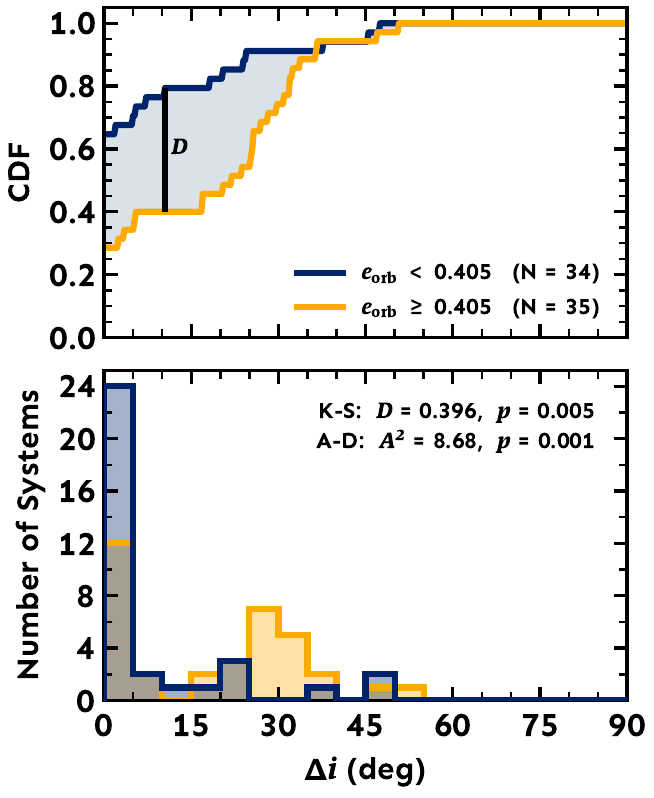}
    \caption{K-S and A-D test results. The population of low-eccentricity orbits (\ecc is $<$ the median, \ecc$_{\rm med}=$0.405) is shown in dark blue, and the high-eccentricity population (\ecc is $\geq$ the median) is shown in gold. \textit{Top}: Cumulative distributions of each population. The vertical black line indicates the K-S statistic, $D$, and the shaded region indicates the discrepant region to which the A-D statistic, $A^2$, is sensitive. \textit{Bottom}: Density histogram of each sample. The results of the K-S and A-D tests and their respective $p$-values are displayed in the upper right corner.}
    \label{fig:KS_AD_Test_Results}
\end{figure}


\section{Results}\label{sec:results}

\subsection{Stellar Obliquity--Eccentricity Correlation}\label{sec:deltai_vs_eccentricity}

Figure~\ref{fig:fig_delta_i_ecc} shows the minimum stellar obliquity, $\Delta i$, as a function of orbital eccentricity for all 69 companions (top) and for the 53 cold Jupiters with companion-to-host mass ratios $M_p/M_\star < 0.02$ (bottom). We find that spin--orbit misalignment is common, and remains so when only cold Jupiters are considered. More importantly, $\Delta i$ depends strongly on orbital eccentricity. Companions on nearly circular orbits ($e < 0.15$) are consistently aligned with their host stars, whereas those on more eccentric orbits span a broad range of $\Delta i$. The dependence of stellar obliquity on eccentricity suggests two distinct dynamical populations: a dynamically cold population with nearly circular, aligned orbits and a dynamically hot population characterized by elevated eccentricities and a wide range of spin--orbit misalignments.  Below, we investigate this potential dichotomy through three complementary statistical analyses.


\begin{itemize}
    \item \textbf{Misalignment Rate:} Companions on nearly circular orbits (\mbox{\ecc$\lesssim$0.1--0.2}) mostly have \Deltai\ close to zero, consistent with spin-orbit alignment. At higher eccentricities, \Deltai\ spans a much broader range from 0\deg to $\sim$50\deg. To determine misalignment rate, we classify each star-planet pair as ``misaligned'' if the MAP of the \Deltai probability distribution departs from 0\deg with $>$95.4\% (2$\sigma$) confidence. With this classification scheme, for the entire sample, we determine that \NMisCirc/\NCirc (\PerMisCirc) companion orbits with \mbox{\ecc$\leq$0.15} are misaligned with the spin of the host star. In contrast, \NMisEcc/\NEcc (\PerMisEcc) companion orbits with \mbox{\ecc$>$0.15} are misaligned. When considering only cold Jupiters, the misalignment rates are 0/16 ($0^{+7}_{-0}\%$) for \mbox{\ecc$<0.15$} and 15/37 (\PerMisEccCJs$\%$) for \mbox{\ecc$\geq 0.15$}.

    \item \textbf{\ecc vs. \Deltai Slope:} We further quantify the dependence of minimum stellar obliquity on concentricity by dividing the sample into five equal-width eccentricity bins and compute the noise-corrected rms of \Deltai in each bin (see Fig~\ref{fig:fig_delta_i_ecc2}). We estimate the intrinsic second moment by subtracting the mean measurement variance from the observed mean squared inclination difference $\Delta i_{\rm rms}=\sqrt{\frac{1}{N}\sum_{j=1}^{N}
            \left(\Delta i_j^2-\sigma_j^2\right)}$,
    where $j$ is the $j$th companion and $N$ is the number of companions in the bin.  We estimate uncertainties using 20,000 bootstrap resamples of the \Deltai posteriors for each companion within each bin, retaining each measurement together with its uncertainties. The error bars span the 16th--84th percentiles of the resulting rms distribution. The \Deltai rms generally increases with eccentricity. For the full sample, the best linear fit gives \Deltai$_{\rm rms}\simeq0.63\,e_{\rm orb}$, with \Deltai$_{\rm rms}$ expressed in radians. 

    \item \textbf{Kolmogorov-Smirnov and Anderson-Darling Test:} To determine whether the low-eccentricity and high-eccentricity orbits are drawn from the same underlying \Deltai distribution, we conduct a Kolmogorov-Smirnov (K-S) test and an Anderson-Darling (A-D) test. We first divide the sample at an eccentricity of $e=0.15$, finding that both tests significantly reject the null hypothesis (K-S $D$=0.54, p=0.0005; A-D $A^{2}$=8.71, p=0.0007) that the two subsamples are drawn from the same distribution. However, this threshold is motivated by the observed pattern and may introduce sensitivity to the choice of eccentricity cut. To assess the robustness of our results and minimize dependence on this choice, we repeat the tests by dividing the full sample at its median eccentricity ($e_{\rm med}=0.405$), yielding two approximately equal-sized subsamples of low-eccentricity ($N=34$) and high-eccentricity ($N=35$) companions.

    Figure \ref{fig:KS_AD_Test_Results} shows the distributions of the two sample and the K-S and A-D statistics. The K-S statistic, $D$, is $\sim$0.40 with $p$-value$=$0.005 indicates a 40\% departure of the two distributions from the null hypothesis with high significance. From the cumulative distributions, we see that roughly 80\% of low-eccentricity companions have \Deltai below $\sim$10\deg compared to $\sim$40\% of high-eccentricity companions. The result of the A-D test provides a complementary comparison. The A-D statistic, $A^2$, yields a value of 8.68---a significant ($p$=0.001) departure from the null hypothesis ($A^2$$=$0). The results of each test shows that the distribution in \Deltai of companions with eccentricities less than the median are not likely to be drawn from the same underlying \Deltai distribution of companions with eccentricities that are higher than the median.

\end{itemize}

\subsection{The Population-Level Distribution of \Deltai}\label{sec:Deltai_distribution}

There is a broad spread in \Deltai across the population of wide-orbit giant companion host stars, ranging from 0\deg to $\sim$50\deg (Figure \ref{fig:fig_delta_i_ecc}). The average \Deltai is \AverageDeltai, and the average precision to which \Deltai is measured is \AverageDeltaiSigma. The breadth of the distribution indicates that many host stars are misaligned with the orbits of cold Jupiters and brown dwarfs. With the method for classifying misalignment outlined in Section \ref{sec:deltai_vs_eccentricity}, across the entire sample, we find that at least \NMisAll/\NTotComp (\PerMisAll) of companion orbits are misaligned with their host star. For cold Jupiters only, the overall misalignment rate is the same with 15/53 (28$\pm$6\%) misaligned systems.

\subsection{\Deltai vs. Other System Properties}\label{sec:Deltai_vs_properties}

\begin{figure*}[tp]
    \centering
    \includegraphics[width=\linewidth]{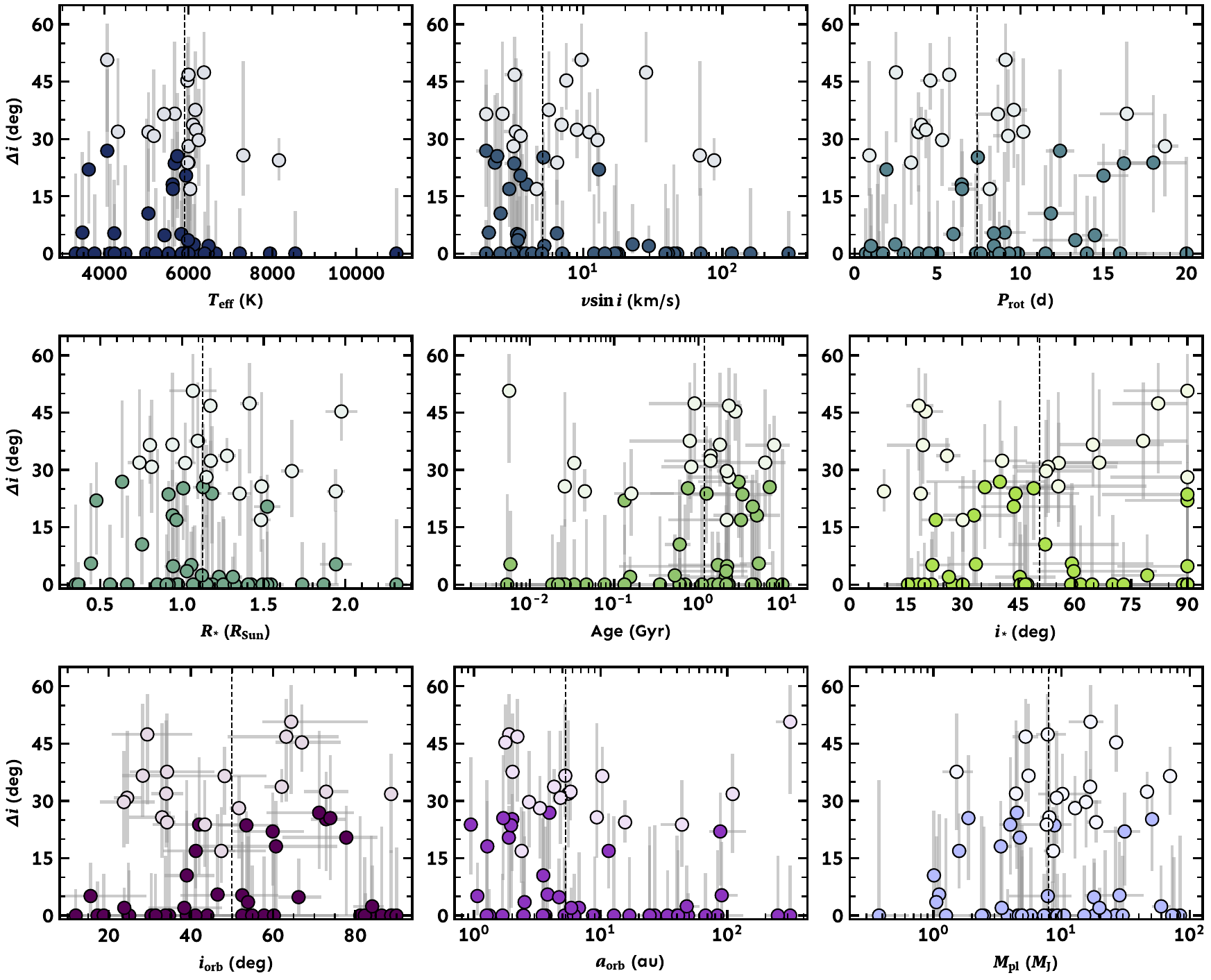}
    \caption{Minimum stellar obliquity, \Deltai, as a function of host star \Teff, \vsini, \Prot, \Rstar, age, \istar, and companion \iorb, \aorb, and \Mpl. Light-colored points are misaligned systems and dark-colored points are systems which show no evidence of misalignment. The median of each parameter is indicated by the vertical dashed line. The results of our analysis comparing \Deltai for systems below and above the median of each parameter (Section \ref{sec:Deltai_vs_properties}) are provided in Table \ref{tab:mosaic_scatter_plots_stats}. This plot and the corresponding statistics determined for planetary mass systems (\Mpl/\Mstar$<$0.02) are displayed in Figure \ref{fig:mosaic_scatter_plots_CJ} and Table \ref{tab:mosaic_scatter_plots_stats_CJ} in the Appendix.} 
    \label{fig:mosaic_scatter_plots}
\end{figure*}

\begingroup
\setlength{\tabcolsep}{3.5pt}
\tabletypesize{\scriptsize}
\begin{deluxetable*}{lcccccccccccc}
\tablecaption{Statistics to test for dependence of \Deltai on host star and companion properties.\label{tab:mosaic_scatter_plots_stats}}
\tablehead{\colhead{Property} & \multicolumn{2}{c}{Pearson Correlation} & \colhead{Median} & \multicolumn{3}{c}{\Deltai Mean (deg)} & \multicolumn{2}{c}{K-S} & \multicolumn{2}{c}{A-D} & \multicolumn{2}{c}{Misalignment Rate} \\[-10.5pt] \multicolumn{1}{c}{} & \multicolumn{2}{c}{\hrulefill} & \multicolumn{1}{c}{} & \multicolumn{3}{c}{\hrulefill} & \multicolumn{2}{c}{\hrulefill} & \multicolumn{2}{c}{\hrulefill} & \multicolumn{2}{c}{\hrulefill} \\[-5.5pt] \colhead{} & \colhead{$r$} & \colhead{$p$} & \colhead{} & \colhead{$<$med} & \colhead{$\geq$med} & \colhead{$\sigma$} & \colhead{$D$} & \colhead{$p$} & \colhead{$A^2$} & \colhead{$p$} & \colhead{$<$med} & \colhead{$\geq$med}}
\startdata
\Teff\ (K) & $-0.05$ & 0.68 & 5941 & $13.6 \pm 15.0$ & $14.4 \pm 16.8$ & 0.04 & 0.15 & 0.820 & $-1.06$ & 0.896 & 6/30 ($20^{+8}_{-7}$\%) & 12/31 ($39^{+9}_{-8}$\%) \\
\vsini\ (km/s) & $-0.20$ & 0.12 & 5.16 & $13.7 \pm 14.3$ & $14.2 \pm 17.4$ & 0.02 & 0.18 & 0.606 & $-0.60$ & 0.586 & 7/30 ($23^{+8}_{-7}$\%) & 11/31 ($35^{+9}_{-8}$\%) \\
\Prot\ (d) & $0.10$ & 0.47 & 7.4 & $13.7 \pm 16.9$ & $15.0 \pm 15.5$ & 0.06 & 0.18 & 0.773 & $-0.43$ & 0.507 & 9/28 ($32^{+9}_{-8}$\%) & 8/28 ($29^{+9}_{-8}$\%) \\
\Rstar\ ($R_\odot$) & $0.02$ & 0.86 & 1.128 & $14.0 \pm 15.5$ & $14.0 \pm 16.4$ & $<$0.01 & 0.15 & 0.812 & $-0.92$ & 0.797 & 7/30 ($23^{+8}_{-7}$\%) & 11/31 ($35^{+9}_{-8}$\%) \\
Age (Gyr) & $0.16$ & 0.21 & 1.249 & $11.3 \pm 16.0$ & $16.6 \pm 15.5$ & 0.24 & 0.28 & 0.155 & $1.88$ & 0.083 & 8/30 ($27^{+8}_{-7}$\%) & 10/31 ($32^{+9}_{-8}$\%) \\
\istar\ (deg) & $-0.10$ & 0.43 & 52 & $14.4 \pm 15.2$ & $13.6 \pm 16.7$ & 0.03 & 0.12 & 0.935 & $-0.80$ & 0.715 & 8/30 ($27^{+8}_{-7}$\%) & 10/31 ($32^{+9}_{-8}$\%) \\
\iorb\ (deg) & $0.02$ & 0.89 & 48.2 & $10.9 \pm 14.4$ & $14.5 \pm 16.5$ & 0.16 & 0.16 & 0.653 & $-0.68$ & 0.613 & 10/34 ($29^{+8}_{-7}$\%) & 9/35 ($26^{+8}_{-7}$\%) \\
\aorb\ (au) & $-0.23$ & 0.06 & 5.57 & $16.7 \pm 15.9$ & $8.9 \pm 14.2$ & 0.36 & 0.33 & 0.032 & $3.96$ & 0.025 & 11/34 ($32^{+8}_{-7}$\%) & 8/35 ($23^{+7}_{-6}$\%) \\
\Mpl\ ($M_{\rm Jup}$) & $0.08$ & 0.51 & 8.0 & $11.9 \pm 15.2$ & $13.6 \pm 15.9$ & 0.08 & 0.14 & 0.845 & $-1.07$ & 0.901 & 7/34 ($21^{+7}_{-6}$\%) & 12/35 ($34^{+8}_{-8}$\%) \\
\ecc & $0.33$ & 0.01 & 0.405 & $7.3 \pm 13.5$ & $18.1 \pm 15.5$ & 0.53 & 0.40 & 0.005 & $8.68$ & 0.001 & 5/34 ($15^{+7}_{-5}$\%) & 14/35 ($40^{+8}_{-8}$\%) \\
\enddata
\tablecomments{$r$ and $p$ are the Pearson correlation coefficient between \Deltai and the corresponding host star or companion property (computed on the logarithm of \vsini, age, \aorb, and \Mpl), and its $p$-value. The sample is split into two subsamples across the median of each property, and Mean$_{\rm <med}$ and Mean$_{\rm \geq med}$ give the mean \Deltai and standard deviation, $s$, of each subsample. $\sigma$ is the significance of the difference between Mean$_{\rm <med}$ and Mean$_{\rm \geq med}$, where $\sigma = |{\rm Mean}_1 - {\rm Mean}_2|/\sqrt{s_1^2 + s_2^2}$. $D$ and $A^2$ are the two-sample Kolmogorov--Smirnov and Anderson--Darling statistics comparing the \Deltai distributions of the two subsamples split across the median. Their respective $p$-values are displayed in the column to the right. The misalignment rates within each subsample are given in the last two columns. Each host star counts once, except in the \iorb, \aorb, \Mpl, and \ecc rows, where each companion counts individually.}
\end{deluxetable*}
\endgroup

Figure \ref{fig:mosaic_scatter_plots} displays \Deltai as a function of stellar effective temperature, \vsini, \Prot, \Rstar, age, \istar, and companion \iorb, \aorb, and \Mpl (we explore \ecc vs. \Deltai in detail in Section \ref{sec:deltai_vs_eccentricity}). We test for a correlation between \Deltai and each respective property by computing the Pearson correlation coefficient, $r$ (computed on the logarithm of \vsini, age, \aorb, and \Mpl).  The coefficient ranges from $r = -0.20$ (\vsini) to $r = 0.16$ (age) with $p>0.05$. These results do not indicate a correlation between \Deltai and any stellar or companion properties listed above.

\Deltai may not follow a linear relationship that the Pearson correlation coefficient specifically tests for, so we also gauge whether \Deltai is systematically concentrated at larger or smaller values if the data were split into distinct subsamples. We split each sample into two subsamples across the median value of each parameter, and compute the \Deltai mean, standard deviation from the mean, and the significance of the difference of the means. In all cases, the means of each subsample differ at the $<$1$\sigma$ level, with the largest discrepancy occurring across the median of \aorb.

For each pair of subsamples, we also conduct K-S and A-D tests (identical to the procedure outlined in Section \ref{sec:deltai_vs_eccentricity}) to determine if \Deltai in each subsample can be drawn from the same underlying distribution. Again, \aorb appears to show the strongest indication that the \Deltai distribution differs between smaller and larger orbital separations. Although the $p$-values of the K-S and A-D tests of \aorb samples can be considered significant by the traditional metric that $p$$<$0.05, compared to those computed for \ecc, they are larger by an order of magnitude.
We further investigated the stellar obliquity distributions and misalignment rates as functions of orbital separation and companion properties in a following work (Zhang $\&$ Biddle, in prep). 

\begin{figure*}[tp]
    \centering
    \includegraphics[width=\linewidth]{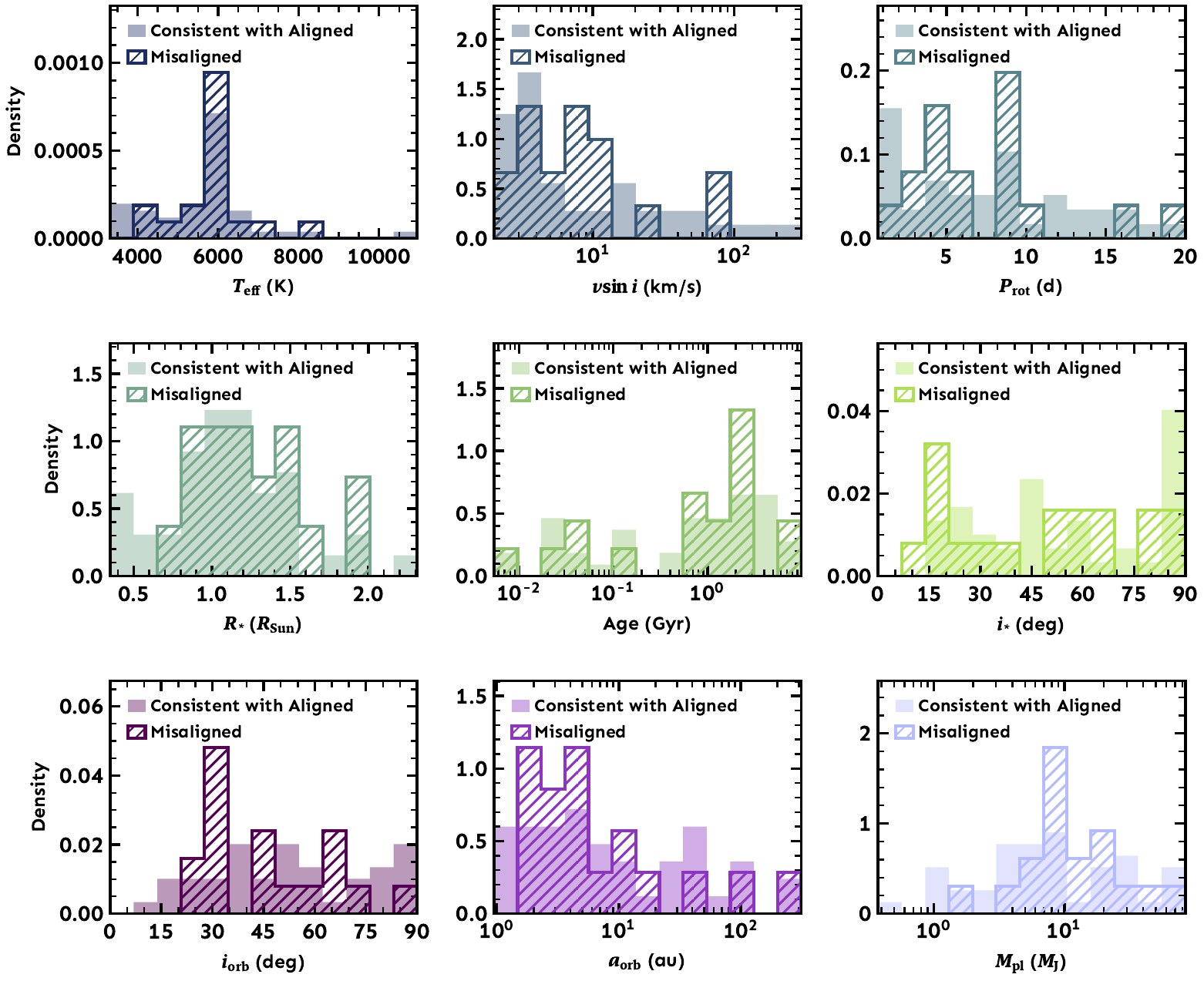}
    \caption{Distributions of system parameters corresponding to systems where \Deltai is misaligned and systems where \Deltai is consistent with alignment.}
    \label{fig:mosaic_histograms}
\end{figure*}

We finally compare the misalignment rates between the subsamples split across the parameter medians. For every parameter except for eccentricity, the rates of misalignment are consistent within 1$\sigma$. For systems with eccentricity less than the median value, we count 5/34 (15$^{+7}_{-5}$\%) misaligned systems. Among companions with eccentricities that are greater than the median, we count 14/35 (40$\pm$8\%) misalignments. The results of each test described above are given in Table \ref{tab:mosaic_scatter_plots_stats}. The distributions of stellar parameters are also similar between  systems  classified as misaligned and those consistent with alignment (Figure \ref{fig:mosaic_histograms}). Overall, we find that stellar obliquity shows the strongest association with orbital eccentricity, while correlations with other stellar and companion properties are either weaker or statistically insignificant.

\subsection{Multi-star Systems vs. Single-star Systems}
\begin{figure}[tp]
    \centering
    \includegraphics[width=0.9\linewidth]{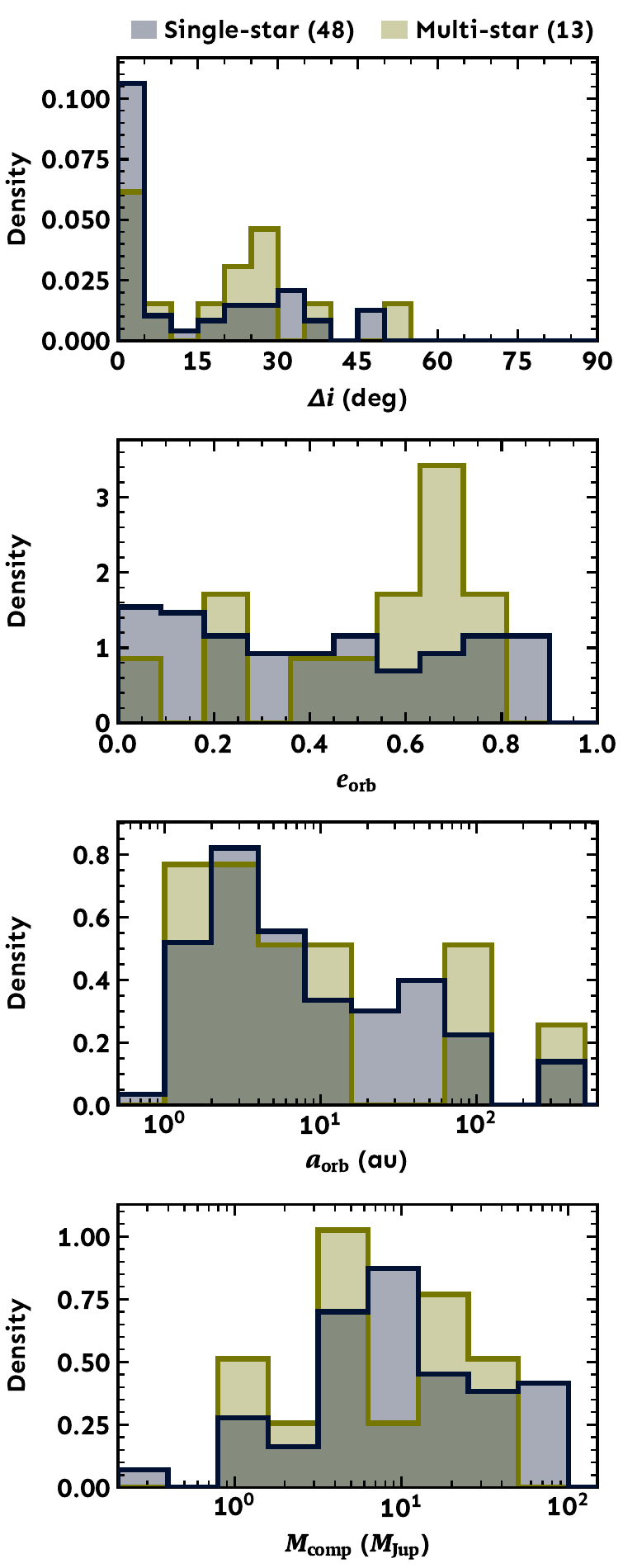}
    \caption{Distributions of \Deltai, \ecc, \aorb, and \Mpl for companions in single-star and multi-star systems.}
    \label{fig:Deltai_eorb_Single_vs_Multi}
\end{figure}

\NMultiHosts of the \NTotStars hosts have at least one bound stellar companion with projected separations spanning \MinCompSep--\MaxCompSep~au with average and median separations of $\sim$2900~au and $\sim$3100~au, respectively. We find that the \Deltai distribution among multi-star systems is consistent with single host stars (Figure \ref{fig:Deltai_eorb_Single_vs_Multi}). The misalignment rates are comparable as well, with \NMisMulti/\NMultiComp (\PerMisMulti) misalignments in multi-star systems, and \NMisSingle/\NSingleComp (\PerMisSingle) misalignments in single-star systems. The distributions of companion masses and orbital separation in each population are also similar. On the other hand, the \ecc distributions of multi-star systems hold a greater share of companions with large-eccentricity orbits compared to single-star systems. We find that 10/34 stars that host companions with orbital eccentricities greater than the median ($e_{\rm med}$$=$0.405) belong to multi-star systems, while 3/27 stars host companions with orbital eccentricities that are less than the median---an odds ratio of 3.3.

If stellar multiplicity is unrelated to orbital eccentricity, the 13 multi-star hosts would be randomly distributed among the \NTotStars stars in the sample, and the number above the median, $k$, would follow a hypergeometric distribution with an expectation value of $\NMultiComp \times 34/\NTotStars \approx 7$. The probability of observing at least 10 multi-star hosts above the median is $P(k \geq 10) = \sum_{k=10}^{13}\binom{34}{k}\binom{27}{13-k}\Big/\binom{\NTotStars}{13} = 0.076$. An excess of 3 systems above the expectation is significant only at a 1.6$\sigma$ level.

\section{Discussion}\label{sec:Discussion}
Our results show that spin-orbit misalignment is common among cold Jupiters and brown dwarf companions beyond 1~au (Section \ref{sec:Deltai_distribution}). Moreover, the misalignment is tied to orbital eccentricity: companions with \ecc$\lesssim$0.15 tend to agree with a spin-orbit aligned orientation, while the dispersion in minimum obliquity increases with larger eccentricity. Here, we discuss potential causes of obliquity and eccentricity excitation in the sample, and place these results into context among star-disk samples and other star-planet samples.

\subsection{Causes of Spin-Orbit Misalignment}\label{sec:causes_of_misalignment}

\subsubsection{Primordial Misalignment}\label{sec:primordial_misalignment}
Spin--orbit misalignment may be inherited from a protoplanetary disk that was already tilted relative to the stellar equator, for example by gravitational torques from a binary companion \citep{Batygin2012}. Recent observational evidence has since shown that misalignment of planet-forming disks occurs even in single-star systems \citep{Biddle2025} that are unlikely to have experienced past stellar fly-bys \citep{Winter2024}. In isolated systems, star--disk misalignment could potentially occur as a result of turbulent formation (e.g., \citealt{Bate2010, Fielding2015}, torques acting on the disk \citep{Lai2014}, or late accretion infall onto the disk from the surrounding environment \citep{Kuffmeier2021, Pelkonen2025, Gupta2026}.

Here, we find that the misalignment rate of cold Jupiters and brown dwarfs---\NMisAll/\NTotComp (\PerMisAll)---is comparable to the primordial misalignment rate observed in protoplanetary disk systems 16/49 (33$^{+7}_{-6}$\%) \citep{Biddle2025}. Besides misalignment rate, the population-level obliquity distribution of our companions is also very
similar to that of the protoplanetary disks in \citet{Biddle2025}. However, the dependence of stellar obliquity on orbital eccentricity in our sample provides additional evidence that is difficult to explain through primordial misalignment alone. We identify two distinct dynamical populations: a dynamically ``cold'' population with nearly circular, well-aligned orbits, and a dynamically ``hot'' population with elevated eccentricities and a broad distribution of stellar obliquities. If planets simply inherited their misalignments from their natal disks, without significant subsequent dynamical excitation, we would also expect to find  circular planets on strongly misaligned orbits. Such systems are notably absent from our sample. The observed eccentricity--obliquity dependence therefore favors a scenario in which substantial spin--orbit misalignments arise primarily through post-formation dynamical interactions that excite both orbital eccentricities and inclinations. Nevertheless, given the typical uncertainties in our minimum obliquity measurements ($\sim15^\circ$--$20^\circ$), we cannot rule out moderate primordial misalignments.  Taken together, our findings suggest that while primordial misalignment may contribute to the observed spin--orbit architectures, post-formation dynamical excitation likely plays an important, and potentially dominant, role in producing the substantial misalignments observed in our sample.


\subsubsection{Planet--planet scattering}\label{sec:planet_planet_scattering} 
Planet--planet scattering has long been a leading explanation for the eccentric orbits of giant exoplanets \citep{RasioFord1996, WeidenschillingMarzari1996}. Giant planets formed in closely packed orbits may become dynamically unstable after disk gas has dispersed \citep{LinIda1997}. The resulting close encounters redistribute orbital energy and angular momentum, causing collisions that generally leave low-eccentricity merger remnants, or ejections  produce more eccentric survivors \citep{Ford2001}.  

The outcome of planet--planet scattering depends on the Safronov number, $\Theta\equiv(v_{\rm esc}/v_{\rm K})^2=2(M_p/M_\star)(a/R_p)$, the squared ratio of a planet's surface escape velocity to its Keplerian orbital velocity \citep{Safronov1972}. When $\Theta \lesssim 1$, collisions limit eccentricity excitation to $e\sim\sqrt{\Theta}$ and generally leave merger remnants on relatively circular, coplanar orbits \citep{Ford2001, Petrovich2014, Anderson2020}. This limit can explain the mass--eccentricity relation of warm Jupiters \citep{Dong2026}. The cold Jupiters and brown dwarf companions in our sample have $\Theta\gg1$, placing them in the ejection-dominated regime, where scattering leaves survivors on eccentric and inclined orbits. Their final eccentricity distribution is relatively insensitive to $\Theta$ and depends mainly on companion mass ratios and the initial orbital architecture \citep{FordRasio2008, JuricTremaine2008}.

Previous $N$-body simulations have shown that planet--planet scattering produces not only eccentric orbits but also inclined ones. \citet{JuricTremaine2008} found that $\gtrsim10\%$ of planets in dynamically relaxed systems end up on orbits inclined by $\gtrsim25^\circ$ to the original plane of the protoplanetary disk. Similarly, \citet{Chatterjee2010} showed that scattering efficiently excites inclinations relative to the initial invariable plane, independent of the initial values. However, these predictions have so far been tested almost exclusively through eccentricities; the inclinations of long-period giant planets relative to their host stars have rarely been measured. Our stellar obliquity measurements provide this missing complementary test. The observed eccentricity--inclination relation has a slope close to $I\sim e/\sqrt{2}$. Dedicated $N$-body simulations are needed to determine whether planet--planet scattering can reproduce this relation. Comparing the predicted and observed joint distributions could constrain the initial planetary architecture, including the number and mass ratios of the planets involved, and test whether a random-kick
description adequately captures the orbital excitation.

\subsubsection{Secular Chaos}\label{sec:secular_chaos}
Secular chaos also can excite eccentricities and inclinations through long-term gravitational interactions among well-spaced planets, without requiring close encounters. Overlapping secular resonances redistribute the system's angular-momentum deficit (AMD) among eccentricity and inclination modes while approximately preserving the planets' semimajor axes \citep{WuLithwick2011, LithwickWu2014}. Unlike planet--planet scattering, this mechanism does not require closely packed orbits or close encounters. Substantial excitation does, however, require sufficient initial AMD, which may be inherited from planet formation, disk interactions, or an earlier scattering phase.

\citet{WuLithwick2011} suggested that secular chaos could explain both the eccentricities and substantial spin--orbit misalignments of giant planets at several au. However, sharing an AMD budget does not necessarily make eccentricity and inclination grow together, so a positive correlation is not a generic prediction. Detailed simulations are therefore needed to determine whether secular chaos can reproduce our observed \ecc--\Deltai relation.

\subsubsection{von Zeipel--Kozai--Lidov cycles}\label{sec:ZKL_cycles}
Von Zeipel--Kozai--Lidov (ZKL) cycles are oscillations in a companion's orbital eccentricity and inclination driven by the gravitational torque of a distant, inclined perturber \citep{1910AN....183..345V, Kozai1962, Lidov1962}. In the classical test-particle, quadrupole limit, large oscillations from an initially nearly circular orbit require a mutual inclination of approximately $39.2^\circ$--$140.8^\circ$ \citep{FabryckyTremaine2007, Naoz2016}. Population studies predict that this mechanism can leave giant planets at several au on moderately eccentric, inclined orbits \citep{Anderson2016, Naoz2016}. However, ZKL cycles do not necessarily produce the observed  \ecc--\Deltai correlation.  As eccentricity and inclination oscillate, companion's orbit can be tilted away from the stellar equator even during a low-eccentricity phase \citep{FabryckyTremaine2007}. Furthermore, the classical ZKL mechanism requires a substantial initial mutual inclination between the companion and the perturber. This mutual inclination may be primordial or generated by earlier dynamical interactions, such as planet--planet scattering. ZKL cycles could therefore act as a subsequent stage of orbital evolution rather than the original source of inclination excitation.



\begin{figure*}[tp]
    \centering
    \includegraphics[width=\linewidth]{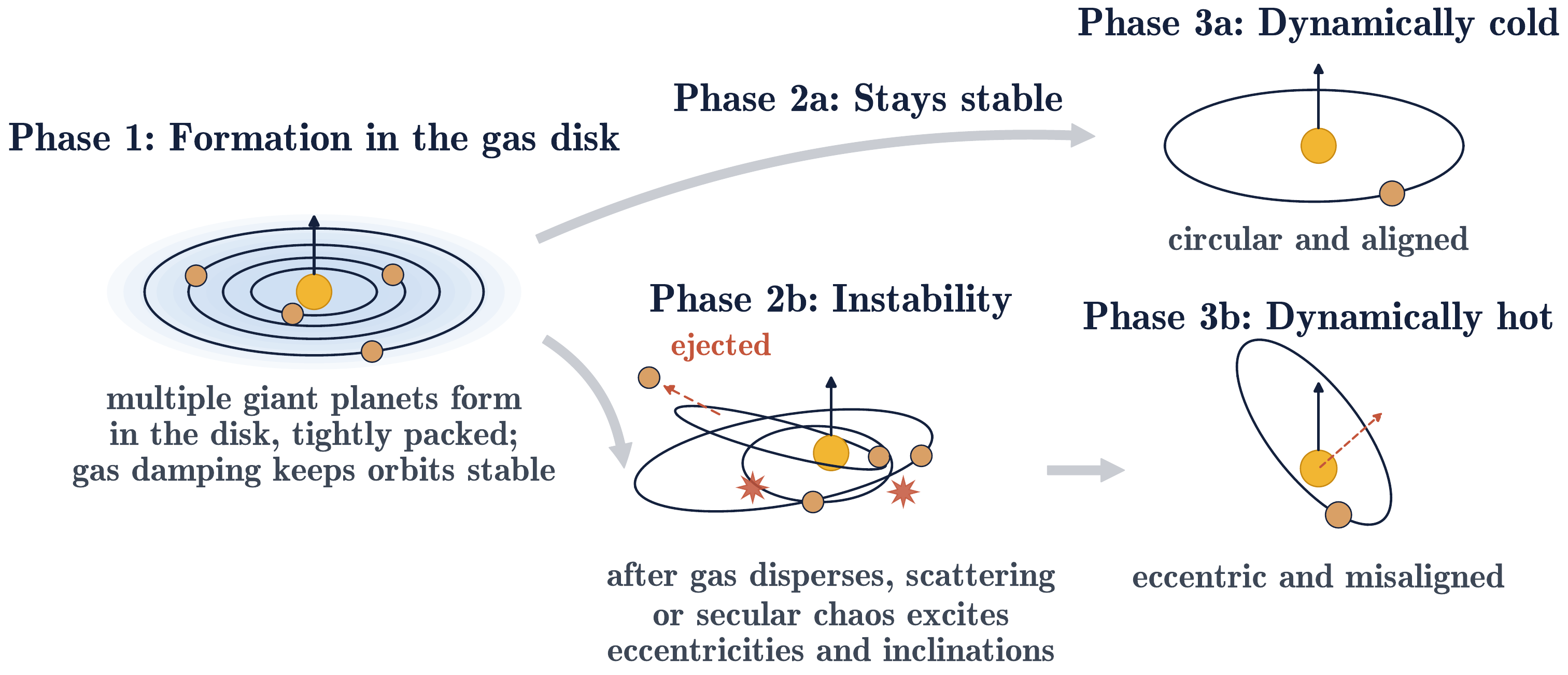}
    \caption{Schematic of two pathways for giant planet evolution. Phase 1: Multiple giant planets form in the protoplanetary disk in tightly packed configuration. Gas damping keeps the orbits stable. Phase 2a-3a: Systems that stay stable after the gas disperses keep their primordial architecture, leaving a dynamically cold population of circular, aligned cold Jupiters. Phase 2b-3b: Systems that become unstable after the gas disperses undergo planet-planet scattering or secular chaos, potentially leading to planetary ejections and leaving surviving planets on dynamically excited, eccentric orbits that are misaligned with their host stars' spin axes.}
    \label{fig:evolution}
\end{figure*}

\subsubsection{A combined picture}
\label{sec:combined_picture}

Figure~\ref{fig:evolution} summarizes the evolutionary pathways suggested by our results. Giant planets form in the protoplanetary disk, often in multi-planet, tightly packed configurations (Phase~1). While the gas disk is present, eccentricity and inclination damping keeps these orbits dynamically stable. After the gas disperses, the systems follow one of two paths. Systems with stable initial setup  retain their primordial architecture (Phase~2a) and form the dynamically cold population of circular, well-aligned companions (Phase~3a). In other systems, the loss of gas damping allows dynamical instabilities and close encounters \citep{RasioFord1996, WeidenschillingMarzari1996, Chatterjee2008, JuricTremaine2008}, or, in more widely spaced systems, slow chaotic exchange of angular momentum among the planets \citep{WuLithwick2011, LithwickWu2014} (Phase~2b). These processes can eject planets and leave the survivors on eccentric, inclined orbits (Phase~3b). Because the stellar spin is decoupled from companions at these separations, it retains its original orientation as the companion's orbit is tilted, and the resulting misalignment is preserved. This produces the dynamically hot population, in which eccentric orbits are accompanied by a broad range of stellar obliquities. Not every excited system becomes strongly misaligned: planet-planet scattering among two planets can raise eccentricity much more efficiently than inclination \citep{Ford2001}, which may explain the eccentric companions in our sample that remain consistent with alignment. If two or more giant planets survive on mutually inclined orbits, their orbital planes subsequently precess around the system's total angular momentum on secular timescales of $\sim10^{4}$--$10^{6}$\,yr. Because the stellar spin remains fixed, the stellar obliquity of each companion oscillates, between $|\theta - \beta|$ and $\theta + \beta$, where $\theta$ and $\beta$ are the inclinations of the orbit and of the stellar spin relative to the invariable plane \citep{Lai2018}.  Since these timescales are much shorter than the system ages, each measured $\Delta i$ is a snapshot at a random phase of this cycle, which broadens the observed obliquity distribution and allows some excited systems to appear aligned at the time of observation.

\subsection{Comparison with Hot and Warm Jupiter Systems}
The eccentricity dependence we find for long-period companions is opposite to that recently reported for close-in giant planets. Using a homogeneous reanalysis of 255 systems with Rossiter--McLaughlin measurements, \citet{WangWangBatygin2026} found that the projected obliquity $\lambda$ of close-in giants depends jointly on eccentricity and planet mass: sub-Saturns can be both eccentric and misaligned, Jupiters are misaligned only on circular orbits, and super-Jupiters and brown dwarfs are aligned at all eccentricities. For Jupiter-mass and more massive close-in companions, eccentricity is therefore anti-correlated with misalignment. In our sample, by contrast, the circular companions are aligned and the misaligned ones are eccentric.

We interpret this contrast as a consequence of the different physical regimes of the two populations rather than of different dynamical origins. Close-in planets have been processed after their orbits were excited. Tides damp eccentricity and obliquity on different timescales, and can circularize an orbit while leaving the star misaligned, or realign the star while some eccentricity remains \citep[see the discussion in][]{WangWangBatygin2026}. The obliquities of hot Jupiters also depend on the host star's effective temperature \citep{Winn2010, Schlaufman2010, Albrecht2012}, a dependence that appears only among circular systems \citep{Rice2022}, pointing to star--planet tidal interaction as an important factor.  A circular, misaligned hot Jupiter may therefore be the end product of high-eccentricity migration that has already been completed, while the eccentric, aligned systems may have followed a different, more nearly coplanar path\citep{zhang2024}. None of these processes operates for companions at several au and beyond, where tides are negligible and the stellar spin is decoupled from the orbit. Our sample thus records the excitation of eccentricity and stellar obliquity as it was produced, before any tidal processing, and this record shows that the two were
raised together.

\subsection{Biases and Systematics}

\subsubsection{Pole-on Host Stars}

The distribution \istar contains 8/\NTotStars stars in pole-on (\istar$<$20\deg) orientations. In a random \istar distribution, the probability of \istar$<$20\deg is $\int_{i_*=0\deg}^{i_*=20^{\circ}} \sin i_*\, di_* = 0.060$, corresponding to an expectation value of about 4 stars in a population of \NTotStars. Applying binomial statistics ($k=8$ out of $n=\NTotStars$ systems, $p=0.060$), the probability of observing at least 8 stars out of \NTotStars with \istar$<$20\deg is $P(k \geq 8\,|\,p=0.060,\,n=\NTotStars) = 1 - P(k<8\,|\,p=0.060,\,n=\NTotStars) = 1 - \sum_{k=0}^{7}\binom{n}{k}p^{k}(1-p)^{(n-k)} = 0.029$. The probability of observing at least 8 pole-on stars in our sample is therefore $\sim$3.0\%. This departure from isotropy is at a $<$2$\sigma$ significance, and with excess of just 4 stars above the expectation, the over-representation of pole-on stars is modest.

\subsubsection{Rotation Period Selection Bias}
Our selection strategy for constructing the sample limits stellar rotation periods to \Prot$\leq$20~d. The decision to omit slowly rotating stars from the final sample likely introduces a selection bias that favors young stars and earlier-type stars, which generally rotate faster than older Sun-like stars \citep{Kraft1967, McQuillan2014}. We see this selection bias emerge in the distributions of age and \Teff in the final sample compared to the parent sample (Figure \ref{fig:HR_Diagram}). In the parent sample, $\sim$88\% of stars have age estimates $>$1 Gyr, while in the final sample, this fraction decreases to $\sim$52\%. Similarly, we see a shift toward hotter \Teff after filtering the parent sample, signifying an increase in the representation of later-type stars compared to the entirety of cold Jupiter and brown dwarf hosts.

\section{Summary}\label{sec:summary}
In this work, we measure the minimum stellar obliquity, \Deltai, for \NTotComp cold Jupiters and brown dwarfs orbiting \NTotStars stars beyond 1~au by comparing angle between stellar rotational inclinations and companion orbital inclinations. We estimate stellar in stellar inclination by combining \Prot, \Rstar, and \vsini, and we determine companion orbital inclinations by jointly fitting radial velocities with \textit{Hipparcos}--\textit{Gaia} absolute astrometry, and, where available, relative astrometry from direct imaging.

The results show that spin--orbit misalignment is common among cold Jupiter and brown dwarf systems, with \Deltai spanning 0\deg--$\sim$50\deg. We find that \NMisAll/\NTotComp (\PerMisAll) companion orbits are misaligned with the spin of their host stars, and that the dispersion in \Deltai appears to increase for orbits with eccentricity greater than $\sim$0.15. In contrast, \Deltai does not appear to show a strong dependence on other stellar or companion properties or the host star's stellar multiplicity. However, there does appear to be a somewhat weaker \Deltai--\aorb dependence, which suggests that companions with wider orbits tend to agree more with aligned orientations, while at closer orbits the spread in \Deltai remains broad. We further investigated this in a following paper (Zhang $\&$ Biddle, in prep).

Comparisons Primordial disk misalignment may contribute to the misaligned population, but it does not by itself excite eccentricity, and ZKL cycles from stellar companions do not generically produce a positive \ecc--\Deltai relation. These results therefore suggest that eccentricity and obliquity were excited together from their initial state, for example through planet--planet scattering or secular chaos. This presents some tension with the primordial obliquity distribution for which approximately 1/3 of protoplanetary disk systems show obliquity misalignment. Dedicated $N$-body and secular simulations that predict the joint distribution of eccentricity and stellar obliquity, together with the expanded sample of astrometric orbits for cold giants expected from \textit{Gaia} DR4, will help determine which post-formation dynamical mechanism drives the orbital architectures of wide-orbit giant companions.

\begin{acknowledgments}

We would like to thank Jerry Xuan, Brendan Bolwer, Josh Winn, Ruth Murray Clay, Dan Fabrycky, and Max Millar-Blanchaer for their helpful scientific discussions. J.Z. would like to acknowledge the use of computational facilities purchased with funds from the National Science Foundation (CNS-1725797) and administered by the Center for Scientific Computing (CSC). The CSC is supported by the California NanoSystems Institute and the Materials Research Science and Engineering Center (MRSEC; NSF DMR 2308708) at UC Santa Barbara. This research has made use of the NASA Exoplanet Archive, which is operated by the California Institute of Technology, under contract with the National Aeronautics and Space Administration under the Exoplanet Exploration Program. This work benefited from the 2025 and 2026 Exoplanet Summer Program in the Other Worlds Laboratory (OWL) at the University of California, Santa Cruz, a program funded by the Heising-Simons Foundation.

\end{acknowledgments}

\begin{contribution}
J.Z. and L.B. contributed equally to this work. J.Z. performed the orbital fitting to derive the companion orbital inclinations, while L.B. derived the stellar inclinations. Both authors contributed equally to the development of the methodology, statistical analysis, interpretation of the results, and manuscript preparation. Both authors reviewed and approved the final manuscript.
\end{contribution}

%
\facilities{HST(STIS), Swift(XRT and UVOT), AAVSO, CTIO:1.3m, CTIO:1.5m, CXO}

\software{astropy \citep{2013A&A...558A..33A,2018AJ....156..123A,2022ApJ...935..167A},  
          Cloudy \citep{2013RMxAA..49..137F}, 
          Source Extractor \citep{1996A&AS..117..393B}
          }


\clearpage
\appendix
\counterwithin{figure}{section}
\counterwithin{table}{section}

\section{Summary of Adopted \vsini}

\begingroup
\small
\setlength{\tabcolsep}{4pt}
\begin{longtable}{lrrl}
\caption{Compiled projected rotational velocities $v\sin{i_*}$ used to derive weighted mean values. The adopted value for each system is shown in bold, with its uncertainty inflated to at least 5\% of the value. Weighted means at or below 2~km~s$^{-1}$ are adopted as upper limits of 2~km~s$^{-1}$. Only the first 3 systems are shown; the full table will be available upon publication.}\label{tab:compiled_vsini}\\
\hline
Host star & $v\sin i$ & Uncertainty & Reference \\
 & (km s$^{-1}$) & (km s$^{-1}$) & \\
\hline
\endfirsthead
\caption[]{Compiled projected rotational velocities $v\sin{i_*}$ used to derive weighted mean values. The adopted value for each system is shown in bold, with its uncertainty inflated to at least 5\% of the value. Weighted means at or below 2~km~s$^{-1}$ are adopted as upper limits of 2~km~s$^{-1}$.}\\
\hline
Host star & $v\sin i$ & Uncertainty & Reference \\
 & (km s$^{-1}$) & (km s$^{-1}$) & \\
\hline
\endhead
\hline
\multicolumn{4}{r}{Continued on next page}\\
\endfoot
\hline
\endlastfoot
16 Cyg B & 3.05 & 2.9 & \cite{vsiniLuck2017} \\
16 Cyg B & 2.45 & 0.13 & \cite{vsiniMaldonado2022} \\
16 Cyg B & 2.3 &  & \cite{vsiniJarvinen2025} \\
16 Cyg B & 2.7 & 1 & \cite{vsiniSoderblom1982} \\
16 Cyg B & 3 &  & \cite{vsiniHoffleit1991} \\
16 Cyg B & 2 &  & \cite{vsiniNordstrom2004} \\
16 Cyg B & 2.2 & 0.5 & \cite{vsiniValenti2005} \\
16 Cyg B & 1.97 &  & \cite{vsiniTakeda2010} \\
16 Cyg B & 1.6 &  & \cite{vsiniBrewer2016} \\
16 Cyg B & 1.4 &  & \cite{vsiniNissen2017} \\
16 Cyg B & 2 &  & \cite{vsiniLubin2024} \\
16 Cyg B & 1.4 & 0.74 & \cite{vsiniFreckelton2025} \\*
16 Cyg B & \textbf{2.40} & \textbf{0.12} & \textbf{Adopted} \\
\hline 
1RXS0342+1216 & 5.04 & 1.51 & \cite{Bowle2023} \\
1RXS0342+1216 & 1.7 &  & \cite{vsiniNewton2016} \\*
1RXS0342+1216 & \textbf{5.04} & \textbf{1.51} & \textbf{Adopted} \\
\hline 
2M0122-2439 & 14.2 & 3.2 & \cite{vsiniMalo2014} \\
2M0122-2439 & 18.1 & 0.5 & \cite{Bowle2023} \\*
2M0122-2439 & \textbf{17.81} & \textbf{0.89} & \textbf{Adopted} \\
\end{longtable}
\endgroup

\begin{figure*}[tp]
    \centering
    \includegraphics[width=0.8\linewidth]{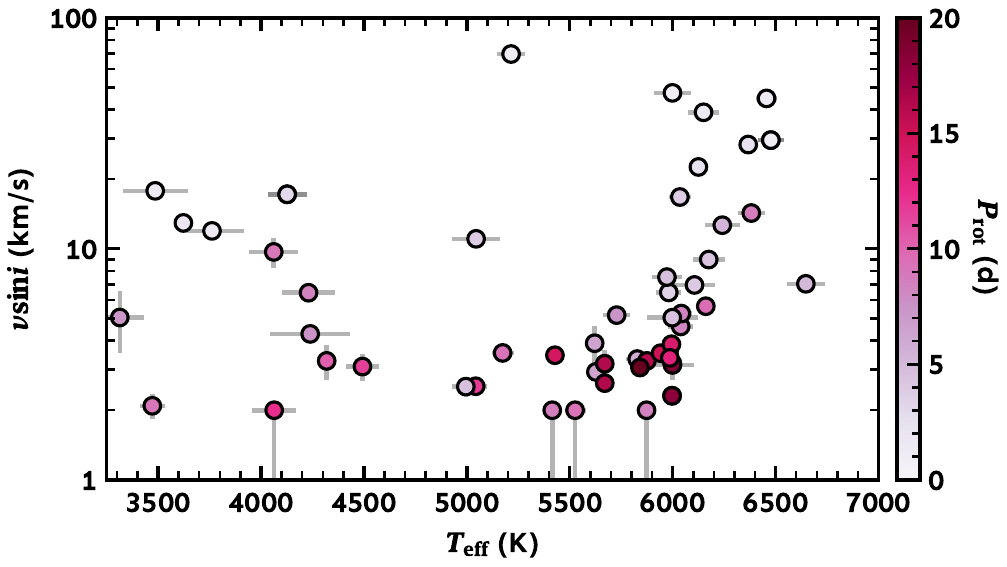}
    \caption{Stellar projected rotational velocity, \vsini, as a function of effective temperature, \Teff. Points are colored according to the stellar rotation period, \Prot.} 
    \label{fig:Teff_vsini_Prot}
\end{figure*}

\clearpage

\onecolumngrid
\section{Materials for rotation period from TESS Light curves}
We retrieve the light curves from the Mikulski Archive for Space Telescopes using \texttt{lightkurve} \citep{Lightkurve2018}. For \textit{TESS}, we download the 2-minute cadence Science Processing Operations Center (SPOC) light curves \citep{Jenkins2016}. For sectors where there are no SPOC data products, we download the TESS-SPOC \citep{Caldwell2020}---both pipelines supply Pre-search Data Conditioning Simple Aperture Photometry (PDCSAP) fluxes, from which instrumental systematics and dilution by neighboring sources have been removed \citep{Stumpe2012, Smith2012, Stumpe2014}. For BD$-$21~0397, we use the TESS light curve from the MIT Quick-Look Pipeline (QLP; \citealt{Huang2020a, Huang2020b}). For \textit{K2} we adopt the PDCSAP light curves, substituting the EVEREST reduction of \citet{luger2016} and \citet{luger2018} where PDCSAP is unavailable or where its six-hour scatter exceeds twice that of EVEREST.

We prepare each light curve by removing non-finite values and normalizing by the median flux. For cases where a star has two or more light curves from consecutive \textit{TESS} sectors, we stitch the light curves together and treat them as one time series. Next, we apply a high-pass Savitsky--Golay filter \citep{savitzky1964} to each series, divide the result from the unfiltered data, and reject points that fall more than three standard deviations from the flattened data.

When determining rotation period from the periodogram results (Section \ref{sec:stellar_rotation_period}), a scenario can arise where star spots reside on opposite stellar hemispheres, causing the strongest peak in the periodogram to coincide with one-half the value of the true rotation period. Therefore, in the event that a light curve produces a strong peak at half of the period identified in one or more other light curves of the same star, we double the the half-period and its uncertainty before computing the weighted mean. All light curves are visually inspected to ensure consistency with the period-search results.

\begin{figure*}[!h]
    \centering
    \includegraphics[width=0.75\linewidth]{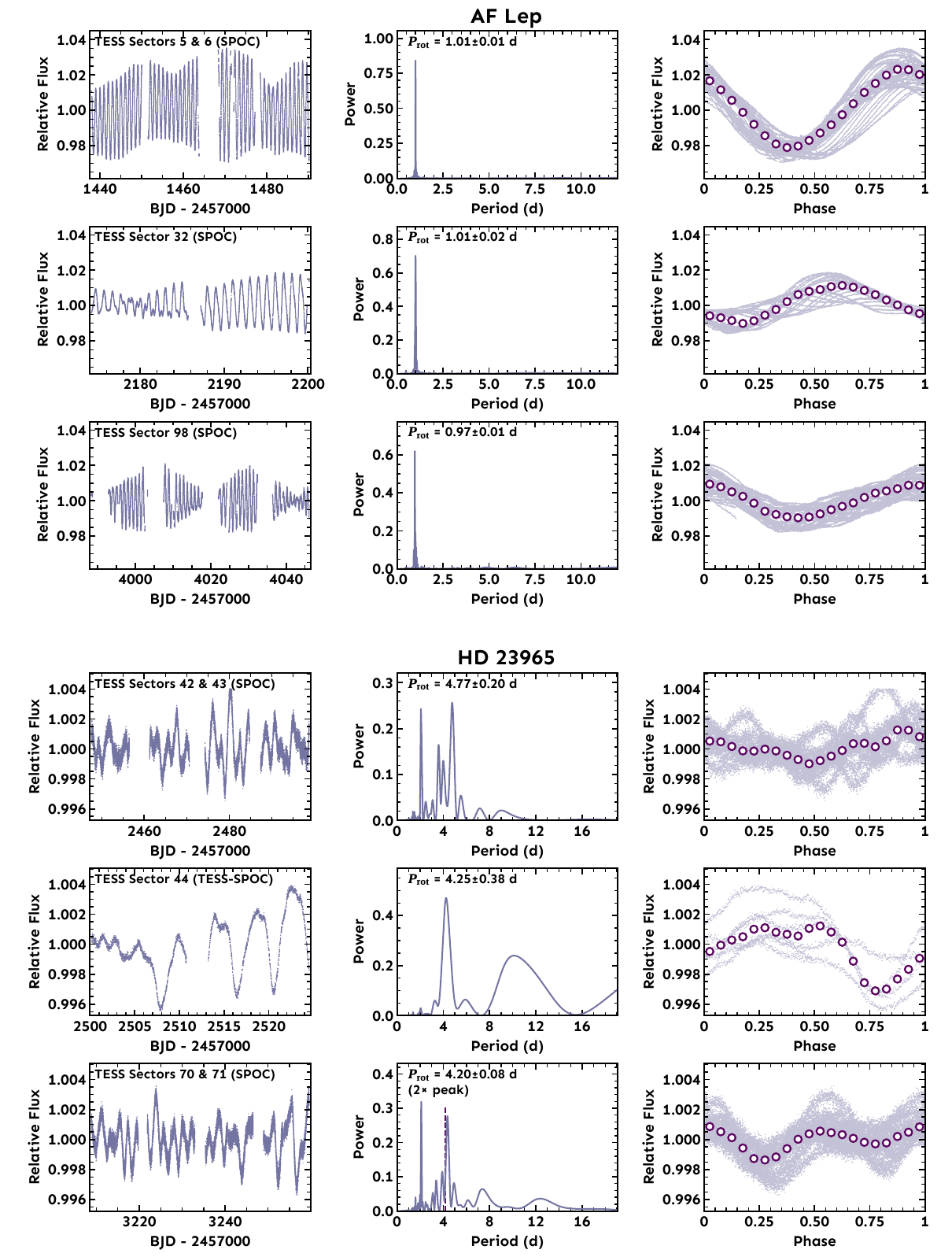}
    \caption{Example period-search results for AF~Lep and HD~223965. The TESS light curves are displayed on the left, the corresponding periodogram is displayed in the center with the adopted period for that printed on the plot face, and the light curve phased to that sector's period is shown on the right. The complete figure set for all systems we have derived rotation period from photometry (40 images) is included in the submission materials and will be made publicly available upon publication of the paper.}
    \label{fig:example_light_curves}
\end{figure*}

\clearpage

\section{Materials for orbital fitting}

\begin{figure*}[!htbp]
    \centering
    \includegraphics[width=\linewidth]{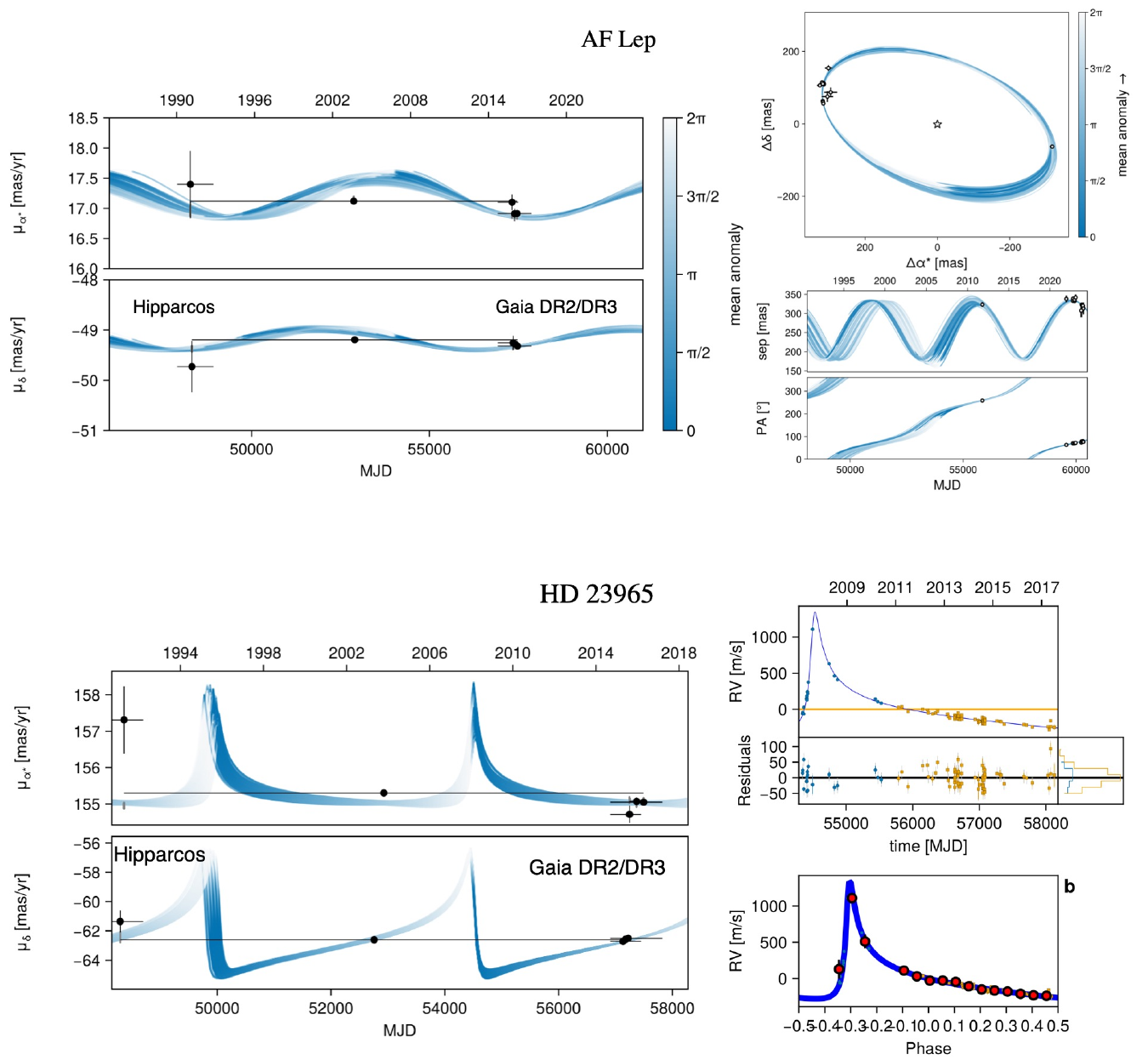}
    \caption{\textit{Top}: 3-D Keplerian orbital models and data of of AF Lep b. 
    Left: observed and fitted absolute astrometry from Hipparcos and Gaia DR2 and DR3 in right ascension and declination. The blue lines are randomly drawn from the posterior distribution. Right: observed and fitted relative astrometry of AF Leb b from direct imaging. \textit{Bottom}: 3-D Keplerian orbital models and data of HD23965 B.  Left: fitted absolute astrometry from Hipparcos and Gaia DR2 and DR3 in right ascension and declination. The blue lines are randomly drawn from the posterior distribution. Right: observed and fitted RVs used in the fitting. The complete figure set for all systems we have done orbital fitting (57 images) is included in the submission materials and will be made publicly available upon publication of the paper. }
    \label{fig:Example_Orbit_Fits}
\end{figure*}

\begin{figure*}[tp]
    \centering
    \includegraphics[width=0.6\linewidth]{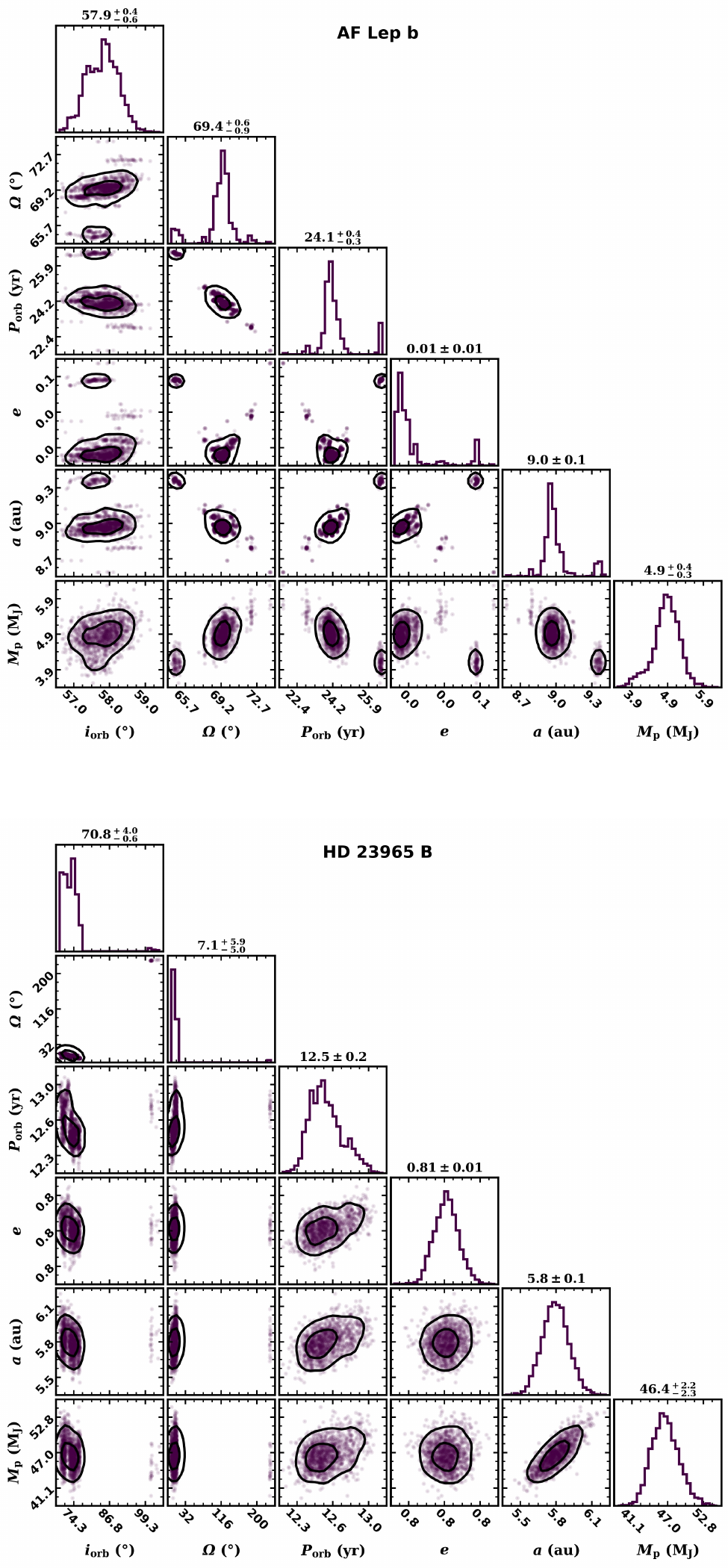}
    \caption{Corner plots showing the orbit fit posteriors of AF~Lep~b and HD~23965~B. The complete figure set for all systems we have done orbital fitting (57 images) is included in the submission materials and will be made publicly available upon publication of the paper.}
    \label{fig:Example_Corner_Plots}
\end{figure*}

\clearpage

\section{Geometry and degeneracy in stellar inclination and orbital inclination}\label{sec:geometry}

\begin{figure}[!b]
    \centering
    \includegraphics[width=0.6\linewidth]{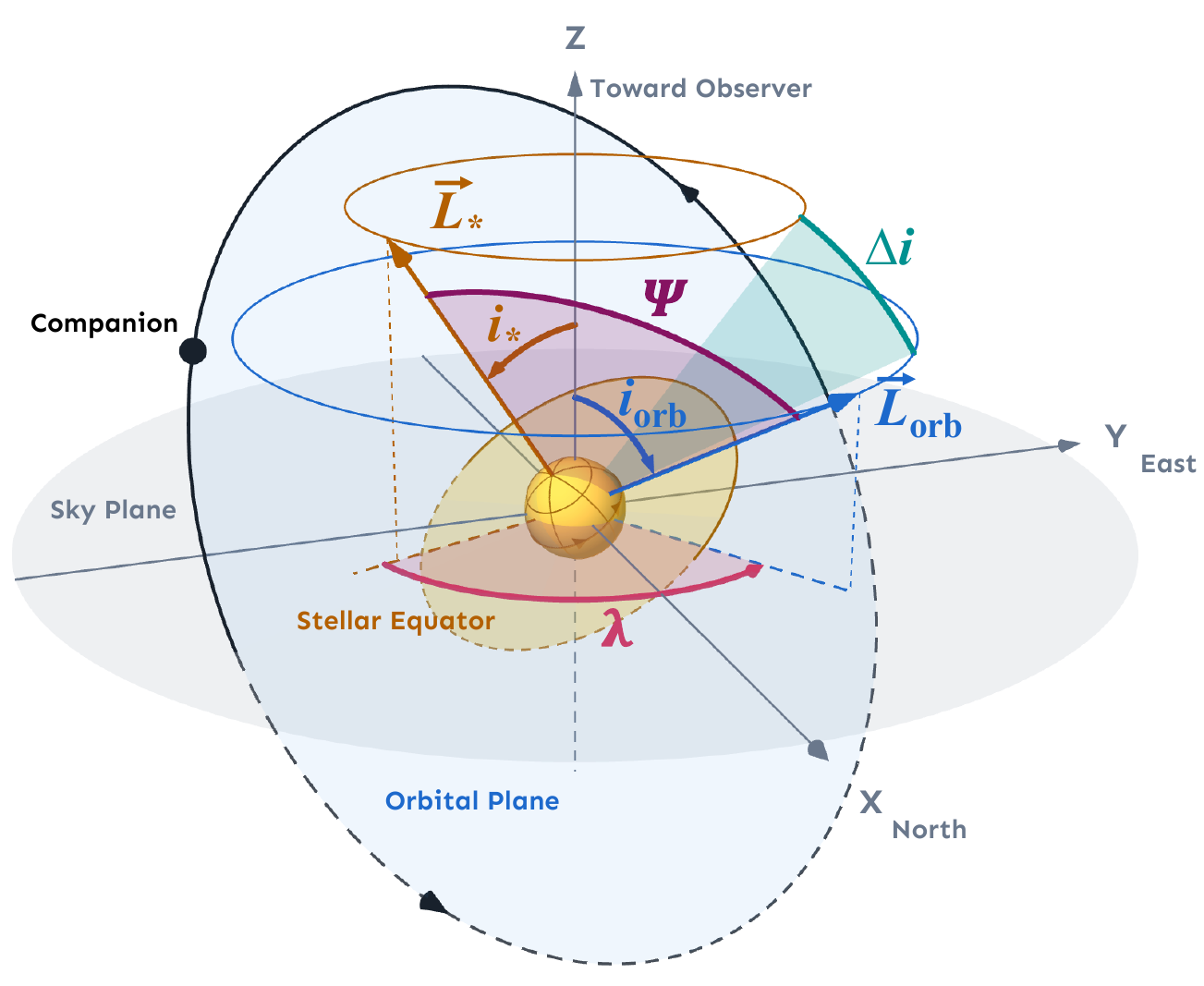}
    \caption{Geometry of stellar obliquity. The angular momentum vectors of the stellar rotation and companion orbit are represented by $\vec{L_*}$ and $\vec{L_{\rm orb}}$, respectively. The inclinations of $\vec{L_*}$ and $\vec{L_{\rm orb}}$ as viewed along the line of sight of the observer (z-axis) are labeled \istar and \iorb. The sky-projected obliquity angle is represented by $\lambda$, and the true three-dimensional angle between $\vec{L_*}$ and $\vec{L_{\rm orb}}$ is the spin--orbit angle, $\Psi$.}
    \label{fig:obliquity_geometry}
\end{figure}

Stellar obliquity, also commonly referred to as spin-orbit angle, is the angle between the stellar spin axis and the normal vector of the orbital plane of planets . The true stellar obliquity angle $\psi$ is given by \citep{FW2009}:
\begin{equation}
\cos\psi = \cos i_\star \cos i_{\rm o} + \sin i_\star \sin i_{\rm o} \cos{\lambda},
\end{equation}
where $i_\star$ and $i_{\rm o}$ are the inclinations of the stellar spin axis and the companion's orbit, and $\lambda$ is the sky-projected stellar obliquity (see Fig.~\ref{fig:obliquity_geometry}). However, while $\lambda$ can be measured through the Rossiter--McLaughlin (RM) effect for transiting planets, it generally remains unknown for non-transiting planets because the position angle of the stellar spin axis on the sky is unconstrained. Consequently, the true stellar obliquity, $\psi$, cannot be directly determined. Following \citet{Bowler2023}, we instead quantify spin--orbit alignment using the minimum stellar obliquity,
\begin{equation}
\Delta i = |i_\star - i_{\rm o}|.
\end{equation}

This observable measures the relative tilt between stellar equatorial plane and companion's orbital plane along the line of sight, and provides a lower limit on the 3-D stellar obliquity.  A nonzero \Deltai means the system is misaligned by at least that amount. A value of \Deltai$\approx0^\circ$ is consistent with alignment but does not rule out misalignment, because the two planes may still differ in $\Omega$. Statistically, however, a small \Deltai generally indicates a small true obliquity $\psi$ (e.g., \citealt{Fabrycky2009, Morton2014}).

\begin{figure*}[tp]
    \centering
    \includegraphics[width=\linewidth]{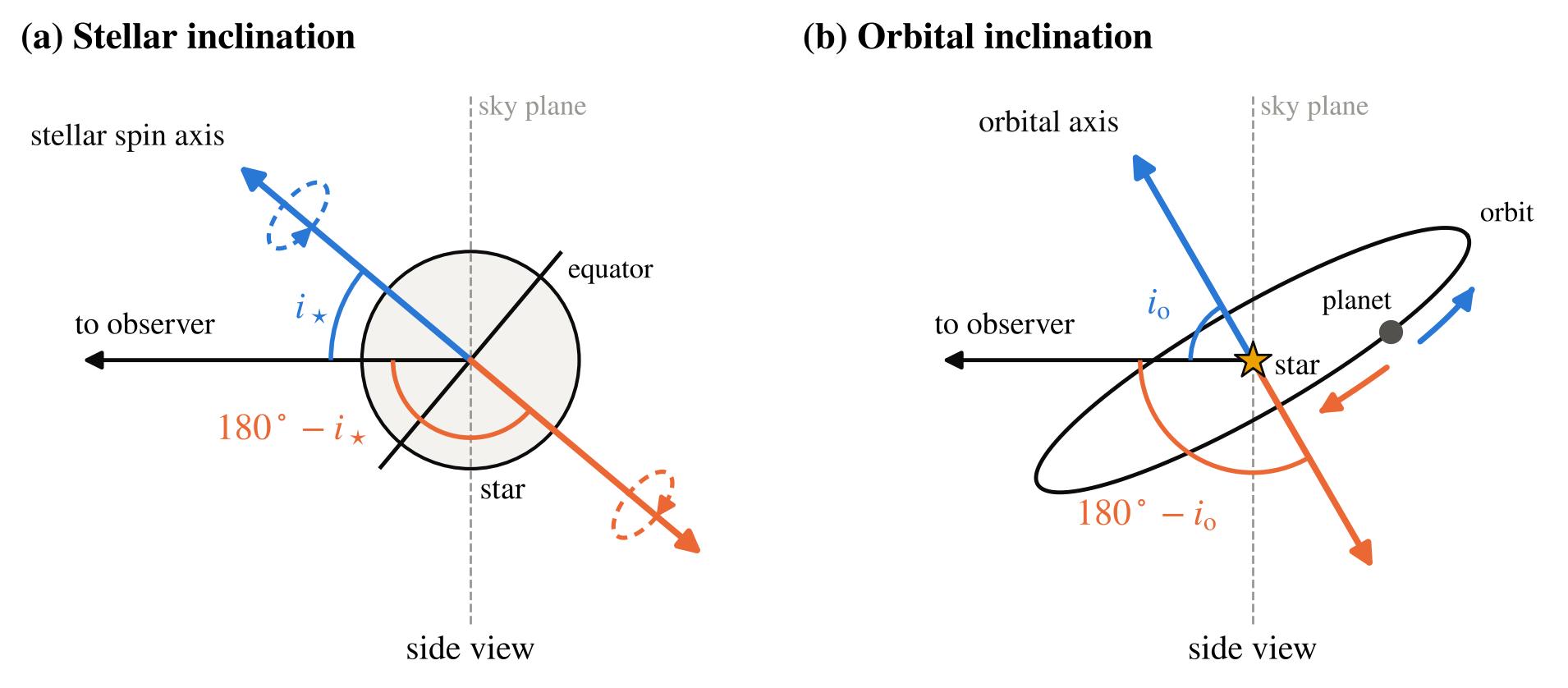}
    \caption{Side views of the prograde--retrograde degeneracies in stellar and orbital inclination. The observer is to the left, and the sky plane is seen edge-on (vertical dashed line). \textbf{(a)} Combining $P_{\rm rot}$, $R_\star$, and $v\sin i_\star$ constrains only $\sin i_\star$, so the stellar spin axis may be inclined by $i_\star$ (blue) or $180^\circ - i_\star$ (orange) relative to the line of sight. The two solutions share the same equatorial plane but have opposite senses of rotation (dashed loops). \textbf{(b)} When the direction of orbital motion on the sky is unconstrained, the orbital axis may likewise be inclined by $i_{\rm o}$ (blue) or $180^\circ - i_{\rm o}$ (orange). Both solutions describe the same orbital plane, traversed in opposite directions (blue and orange arrows next to the planet). Because we are interested in the angle between the two planes rather than the sense of rotation, we restrict both $i_\star$ and $i_{\rm o}$ to $0^\circ$--$90^\circ$.}
    \label{fig:fig_sideview}
\end{figure*}

Interpreting \Deltai is further complicated by the pro-grade and retrograde degeneracy of both stellar inclination and planet orbits (see Fig.~\ref{fig:fig_sideview}). First, \vsini and \Prot constrain only $\sin i_\star$, so they carry no information about the direction of stellar rotation. The spin axis may be inclined by \istar or by 180\deg$-$\istar; both solutions describe the same equatorial plane, but with opposite angular momentum vectors. Second, for some companions the orbit-fitting posterior of \iorb is bimodal, with peaks at \iorb and 180\deg$-$\iorb. These two solutions likewise describe the same orbital plane, traversed in opposite directions, and correspond to the usual prograde--retrograde degeneracy. For each companion, we calculated the posterior distribution of the inclination
difference, \Deltai~$=\left|i_{\rm p}-i_\star\right|$, using Monte Carlo
sampling. We drew $10^5$ independent samples from the posterior distributions
of the orbital inclination, \iorb, and stellar inclination, \istar. We adopted the maximum-a-posteriori value of this
density as the nominal \Deltai. A contiguous 68$\%$ highest-density interval
was constructed around the posterior mode, and the reported lower and upper
uncertainties correspond to the differences between the mode and the lower and
upper bounds of this interval, respectively.

\clearpage

\section{\Deltai as a function of stellar and orbital properties of cold Jupiters only}

\begin{figure*}[!h]
    \centering
    \includegraphics[width=\linewidth]{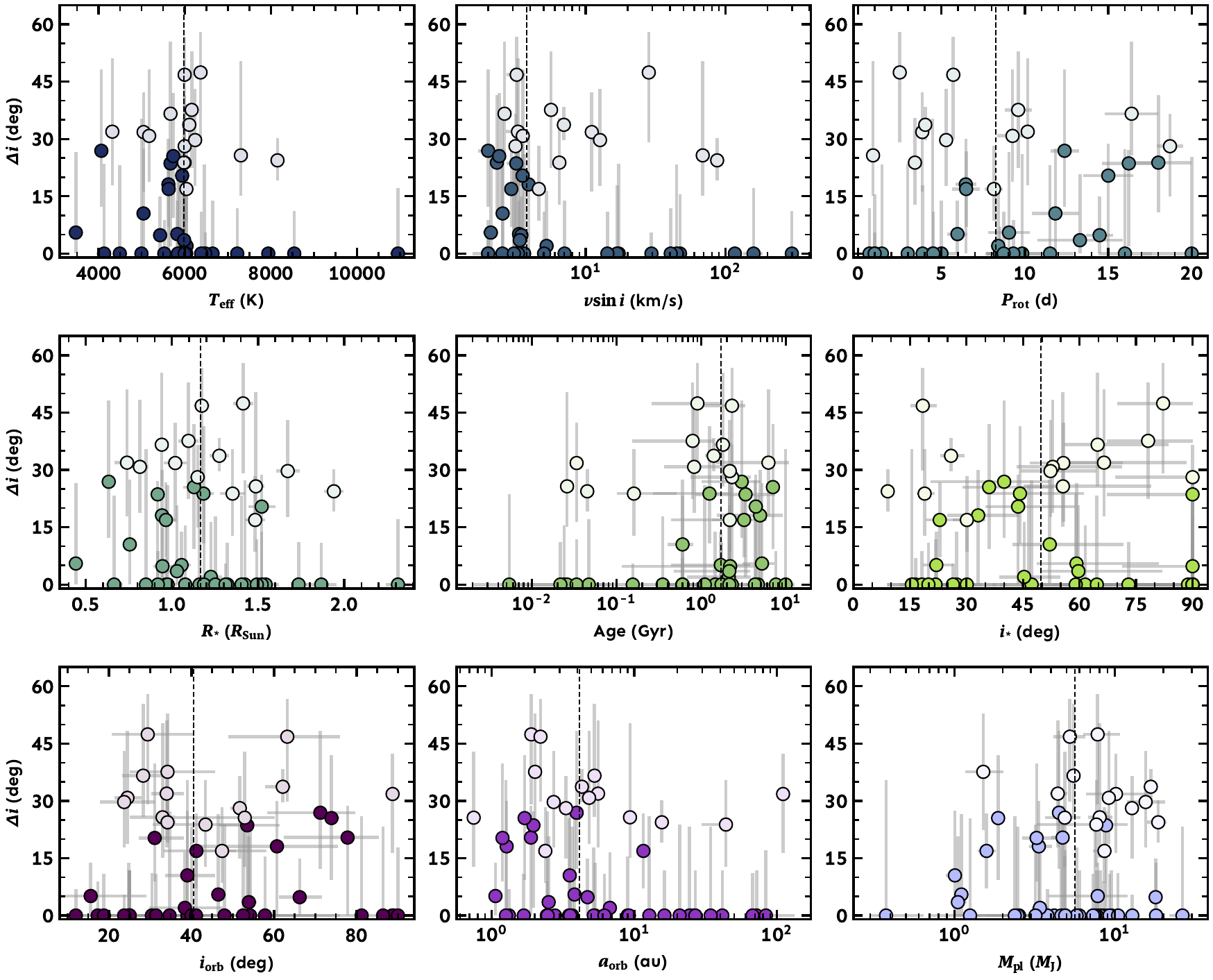}
    \caption{Same as Figure \ref{fig:mosaic_scatter_plots}, but for systems with \Mpl/\Mstar$<$0.02.} 
    \label{fig:mosaic_scatter_plots_CJ}
\end{figure*}

\begingroup
\setlength{\tabcolsep}{3.5pt}
\tabletypesize{\scriptsize}
\begin{deluxetable*}{lcccccccccccc}
\tablecaption{Statistics to test for dependence of \Deltai on host star and companion properties for cold Jupiter systems ($M_{\rm p}/M_* < 0.02$) only.\label{tab:mosaic_scatter_plots_stats_CJ}}
\tablehead{\colhead{Property} & \multicolumn{2}{c}{Pearson Correlation} & \colhead{Median} & \multicolumn{3}{c}{\Deltai Mean (deg)} & \multicolumn{2}{c}{K-S} & \multicolumn{2}{c}{A-D} & \multicolumn{2}{c}{Misalignment Rate} \\[-10.5pt] \multicolumn{1}{c}{} & \multicolumn{2}{c}{\hrulefill} & \multicolumn{1}{c}{} & \multicolumn{3}{c}{\hrulefill} & \multicolumn{2}{c}{\hrulefill} & \multicolumn{2}{c}{\hrulefill} & \multicolumn{2}{c}{\hrulefill} \\[-5.5pt] \colhead{} & \colhead{$r$} & \colhead{$p$} & \colhead{} & \colhead{$<$med} & \colhead{$\geq$med} & \colhead{$\sigma$} & \colhead{$D$} & \colhead{$p$} & \colhead{$A^2$} & \colhead{$p$} & \colhead{$<$med} & \colhead{$\geq$med}}
\startdata
\Teff\ (K) & $-0.14$ & 0.36 & 5984 & $13.6 \pm 13.1$ & $13.9 \pm 17.0$ & 0.01 & 0.22 & 0.660 & $-0.32$ & 0.430 & 5/23 ($22^{+9}_{-8}$\%) & 9/23 ($39^{+10}_{-9}$\%) \\
\vsini\ (km/s) & $-0.18$ & 0.23 & 3.8 & $14.8 \pm 14.5$ & $12.7 \pm 15.7$ & 0.10 & 0.26 & 0.422 & $0.21$ & 0.305 & 5/23 ($22^{+9}_{-8}$\%) & 9/23 ($39^{+10}_{-9}$\%) \\
\Prot\ (d) & $0.08$ & 0.62 & 8.4 & $14.8 \pm 16.6$ & $13.6 \pm 14.2$ & 0.05 & 0.12 & 0.993 & $-0.88$ & 0.769 & 8/20 ($40^{+11}_{-10}$\%) & 5/21 ($24^{+10}_{-8}$\%) \\
\Rstar\ ($R_\odot$) & $-0.11$ & 0.48 & 1.17 & $14.7 \pm 14.0$ & $12.8 \pm 16.2$ & 0.09 & 0.26 & 0.422 & $0.39$ & 0.265 & 6/23 ($26^{+9}_{-8}$\%) & 8/23 ($35^{+10}_{-9}$\%) \\
Age (Gyr) & $0.15$ & 0.31 & 1.71 & $12.8 \pm 15.8$ & $14.7 \pm 14.4$ & 0.09 & 0.22 & 0.660 & $0.13$ & 0.316 & 8/23 ($35^{+10}_{-9}$\%) & 6/23 ($26^{+9}_{-8}$\%) \\
\istar\ (deg) & $0.01$ & 0.96 & 50 & $12.4 \pm 13.9$ & $15.1 \pm 16.2$ & 0.13 & 0.26 & 0.422 & $-0.43$ & 0.489 & 5/23 ($22^{+9}_{-8}$\%) & 9/23 ($39^{+10}_{-9}$\%) \\
\iorb\ (deg) & $0.02$ & 0.87 & 41 & $11.6 \pm 15.7$ & $13.0 \pm 14.0$ & 0.07 & 0.17 & 0.759 & $-0.42$ & 0.478 & 8/26 ($31^{+9}_{-8}$\%) & 7/27 ($26^{+9}_{-8}$\%) \\
\aorb\ (au) & $-0.26$ & 0.06 & 4.296 & $15.1 \pm 15.3$ & $9.7 \pm 13.9$ & 0.26 & 0.28 & 0.189 & $1.57$ & 0.110 & 7/26 ($27^{+9}_{-8}$\%) & 8/27 ($30^{+9}_{-8}$\%) \\
\Mpl\ ($M_{\rm Jup}$) & $0.12$ & 0.41 & 5.77 & $12.6 \pm 14.8$ & $12.1 \pm 14.9$ & 0.03 & 0.10 & 0.994 & $-0.88$ & 0.750 & 5/26 ($19^{+8}_{-7}$\%) & 10/27 ($37^{+9}_{-9}$\%) \\
\ecc & $0.38$ & $<$0.01 & 0.336 & $5.8 \pm 12.5$ & $18.6 \pm 14.1$ & 0.68 & 0.51 & 0.001 & $9.81$ & 0.001 & 3/26 ($12^{+7}_{-5}$\%) & 12/27 ($44^{+9}_{-9}$\%) \\
\enddata
\end{deluxetable*}
\endgroup

\clearpage

\onecolumngrid
\section{Summary of stellar parameters, orbital fitting results and stellar obliquities}
\begingroup
\fontsize{6}{7}\selectfont
\begin{center}
\begin{longtable*}{lcccccccc}
\caption{Stellar parameters for our sample. Only the first 10 rows are shown; the full table will be available upon publication.}\label{tab:star_parameters}\\
\hline
\shortstack{Host\\star\\name} & \shortstack{Sp\\Type} & \shortstack{$T_{\rm eff}$\\(K)} & \shortstack{$R_\star$\\($R_\odot$)} & \shortstack{$M_\star$\\($M_\odot$)} & \shortstack{$v\sin i$\\(km s$^{-1}$)} & \shortstack{$P_{\rm rot}$\\(d)} & \shortstack{$G$\\(mag)} & \shortstack{Age\\(Gyr)} \\
\hline
\endfirsthead
\caption[]{Stellar parameters for our sample (continued).}\\
\hline
\shortstack{Host\\star\\name} & \shortstack{Sp\\Type} & \shortstack{$T_{\rm eff}$\\(K)} & \shortstack{$R_\star$\\($R_\odot$)} & \shortstack{$M_\star$\\($M_\odot$)} & \shortstack{$v\sin i$\\(km s$^{-1}$)} & \shortstack{$P_{\rm rot}$\\(d)} & \shortstack{$G$\\(mag)} & \shortstack{Age\\(Gyr)} \\
\hline
\endhead
\hline
\multicolumn{9}{r}{Continued on next page}\\
\endfoot
\hline
\multicolumn{9}{l}{\textit{1.} $T_{\rm eff}$, $M_\star$, $R_\star$, and spectral types are from the TESS Input Catalog \citep{Tessinput2019}.}\\
\multicolumn{9}{l}{\textit{2.} $G$ magnitudes are from \cite{Gaiadr3}.}\\
\multicolumn{9}{l}{\parbox[t]{0.95\textwidth}{%
    \raggedright
\textit{3.}  The $v\sin{i_*}$ values listed here are the weighted means adopted in our analysis. A compilation of individual $v\sin{i_*}$ measurements from the literature and their references is provided in Table~\ref{tab:compiled_vsini}.}%
}\\
\multicolumn{9}{l}{\parbox[t]{0.95\textwidth}{%
    \raggedright
\textit{4.} Sources of the adopted rotation periods:
51 Eri from \citep{Sepulveda2022}; 
BD-210397 from \citep{Lin2024};
GQ Lup from \citep{Donati2012}; 
HD 147513 from \citep{Cincunegui2007}; 
HD 196050, HD 213240, and HD 220689 from \citep{TejadaArevalo2021}; 
$\pi$ Men from \citep{Hatzes2022}; 
HD 75302 from \citep{Wright2011}; 
HR 7672 from \citep{Donahue1996}; 
HR 8799 from \citep{Sepulveda2023}; 
HD 82943 from \citep{Mayor2004}; 
HD 206893 from \citep{Delorme2017};
ROXs 12 from \citep{Bowler2017}; 
Kepler-419 from \citep{Angus2018};
HD 7449 from \citep{Dumusque2011}.
All other listed rotation periods are measured from TESS/K2 photometry in this work (see Appendix A).}%
}\\
\endlastfoot
1RXS0342+1216 & M5.2 & $3314\pm 119$ & $0.35\pm 0.01$ & $0.33\pm 0.02$ & $5.04\pm 1.51$ & $7.32\pm 0.18$ & 13.24 & $0.695\pm 0.085$ \\
2M0122-2439 & M3.5 & $3486\pm 157$ & $0.37\pm 0.01$ & $0.35\pm 0.02$ & $17.81\pm 0.87$ & $1.49\pm 0.01$ & 12.9 & $0.133\pm 0.015$ \\
2M2351+3127 & M2 & $3623\pm 7$ & $0.47\pm 0.01$ & $0.47\pm 0.02$ & $12.94\pm 0.59$ & $1.93\pm 0.04$ & 12.51 & $0.133\pm 0.015$ \\
51 Eri & F0I & $7303\pm 78$ & $1.49\pm 0.05$ & $1.65\pm 0.27$ & $68.78\pm 1.63$ & $0.9\pm 0.2$ & 5.16 & $0.026\pm 0.003$ \\
AB Pic & K1 & $5045\pm 116$ & $1.02\pm 0.07$ & $0.84\pm 0.1$ & $11.04\pm 0.29$ & $3.85\pm 0.06$ & 8.84 & $0.0337\pm 0.002$ \\
AF Lep & F6 & $5999\pm 90$ & $1.08\pm 0.04$ & $1.29\pm 0.2$ & $47.29\pm 1.11$ & $1.01\pm 0.01$ & 6.19 & $0.024\pm 0.003$ \\
BD-210397 & K7 & $4319\pm 5$ & $0.74\pm 0.07$ & $0.64\pm 0.08$ & $3.27\pm 0.58$ & $10.17\pm 0.12$ & 9.26 & $6.2\pm 4.7$ \\
beta Pic & A6 & $7940\pm 24$ & $1.54\pm 0.08$ & 1.95 & $116.82\pm 2.75$ &  & 3.82 & $0.026\pm 0.003$ \\
eps Eri & K2 & $5043\pm 53$ & $0.76\pm 0.04$ & $0.85\pm 0.1$ & $2.54\pm 0.11$ & $11.83\pm 1.4$ & 3.48 & $0.6\pm 0.2$ \\
GJ 504 & G0 & $5981\pm 61$ & $1.35\pm 0.07$ & $1.12\pm 0.15$ & $6.47\pm 0.13$ & $3.41\pm 0.17$ & 5.04 & $0.16^{+0.35}_{-0.06}$ \\
\end{longtable*}
\end{center}
\endgroup

\begingroup
\fontsize{6}{7}\selectfont
\setlength{\tabcolsep}{3pt}
\begin{center}
\begin{longtable*}{llcccc}
\caption{Orbital fit parameters for companions in our sample. Only the first 10 rows are shown; the full table will be available upon publication.}\label{tab:orbit_fit}\\
\hline
Name & Planet & $i_{\rm orb}$ & $e$ & $a$ & $M_{\rm pl}$ \\
 & Letter & deg &  & AU & $M_{\rm Jup}$ \\
\hline
\endfirsthead
\caption[]{Orbital fit parameters for companions in our sample (continued).}\\
\hline
Name & Planet & $i_{\rm orb}$ & $e$ & $a$ & $M_{\rm pl}$ \\
 & Letter & deg &  & AU & $M_{\rm Jup}$ \\
\hline
\endhead
\hline
\multicolumn{6}{r}{Continued on next page}\\
\endfoot
\hline
\endlastfoot
1RXS0342+1216 & B & $84.2^{+0.8}_{-1.0}$ & $0.3^{+0.3}_{-0.0}$ & $40.1^{+2.0}_{-6.6}$ & $84.0^{+10.5}_{-9.2}$ \\
2M0122-2439 & b & $97.5^{+2.0}_{-1.9}$ & $0.6^{+0.0}_{-0.4}$ & $49.0^{+11.1}_{-11.4}$ & $26.5^{+7.1}_{-12.7}$ \\
2M2351+3127 & B & $121.0^{+5.7}_{-8.6}$ & $0.6^{+0.1}_{-0.3}$ & $88.0^{+52.7}_{-15.7}$ & $31.1^{+9.8}_{-9.8}$ \\
51 Eri & b & $146.4^{+8.6}_{-1.7}$ & $0.6^{+0.0}_{-0.0}$ & $9.3^{+0.8}_{-0.1}$ & $8.0^{+1.6}_{-1.2}$ \\
AB Pic & b & $90.8^{+1.0}_{-0.3}$ & $0.5^{+0.1}_{-0.0}$ & $110.3^{+3.7}_{-10.2}$ & $10.1^{+6.1}_{-3.5}$ \\
AF Lep & b & $57.9^{+0.4}_{-0.5}$ & $0.0^{+0.0}_{-0.0}$ & $9.0^{+0.1}_{-0.0}$ & $4.9^{+0.4}_{-0.3}$ \\
BD-210397 & c & $34.2^{+3.8}_{-3.5}$ / $146.3^{+3.7}_{-3.1}$ & $0.4^{+0.0}_{-0.0}$ & $5.6^{+0.2}_{-0.2}$ & $4.4^{+0.7}_{-0.5}$ \\
beta Pic & c & $88.5^{+0.5}_{-0.5}$ & $0.1^{+0.1}_{-0.0}$ & $2.8^{+0.0}_{-0.0}$ & $8.9^{+1.5}_{-1.7}$ \\
beta Pic & b & $89.0^{+0.1}_{-0.0}$ & $0.0^{+0.1}_{-0.0}$ & $9.2^{+1.0}_{-0.1}$ & $4.2^{+1.3}_{-2.8}$ \\
beta Pic & d & $88.7^{+0.9}_{-0.3}$ & $0.1^{+0.1}_{-0.1}$ & $25.9^{+2.2}_{-3.4}$ & $2.5^{+2.3}_{-1.3}$ \\
\end{longtable*}
\end{center}
\endgroup

\begingroup
\fontsize{6}{7}\selectfont
\setlength{\tabcolsep}{3pt}
\setlength{\LTpost}{3pt}
\begin{longtable*}{llcccc}
\caption{Stellar inclinations, folded orbital inclinations, and $\Delta i$, with its $95.4\%$ credible interval. Only the first 10 systems are shown; the full table will be available upon publication.}\label{tab:inclinations_delta_i}\\
\hline
Host star & Planet & $i_\star$ & $i_{\rm orb,folded}$ & $\Delta i$ & $95.4\%$ CI for $\Delta i$ \\
 & Letter & (deg) & (deg) & (deg) & (deg) \\
\hline
\endfirsthead
\caption[]{Stellar inclinations, folded orbital inclinations, and $\Delta i$ (continued).}\\
\hline
Host star & Planet & $i_\star$ & $i_{\rm orb,folded}$ & $\Delta i$ & $95.4\%$ CI for $\Delta i$ \\
 & Letter & (deg) & (deg) & (deg) & (deg) \\
\hline
\endhead
\hline
\multicolumn{6}{r}{Continued on next page}\\
\endfoot
\multicolumn{6}{l}{\parbox[t]{0.7\textwidth}{%
    \raggedright
\textit{1.} The orbital inclinations of HR 8799 b, c, d, and e are adopted from \cite{zurlo2016}; those of PDS 70 b and c, TOI-4562 c, and Kepler-419 c are adopted from\cite{Wang2021PDS70}, \cite{Fermiano2024}, and \cite{Dawson2014}, respectively. All other orbital inclinations are derived from orbital fitting in this work. Only the first 10 systems are shown; the full table will be available upon publication.}%
}\\
\multicolumn{6}{l}{\parbox[t]{0.7\textwidth}{%
    \raggedright
\textit{2.} The stellar inclinations of $\beta$ Pic, 16 Cyg B, HIP 99770, $\kappa$ And, and HIP 65426 are adopted from \cite{Zwintz2019}, \cite{Davies2015}, \cite{Balmer2026},\cite{Jones2016}, and \cite{Sepulveda2024}, respectively. All other stellar inclinations are derived in this work.}%
}\\
\endlastfoot
1RXS0342+1216 & B & $90.00^{+0.00}_{-27.70}$ & $83.90^{+1.58}_{-1.69}$ & $0.00^{+20.88}_{-0.00}$ & $[0.0,\,46.5]$ \\
2M0122-2439 & b & $90.00^{+0.00}_{-8.60}$ & $82.40^{+2.13}_{-2.30}$ & $0.00^{+5.84}_{-0.00}$ & $[0.0,\,10.6]$ \\
2M2351+3127 & B & $90.00^{+0.00}_{-16.70}$ & $59.90^{+7.56}_{-9.63}$ & $22.00^{+10.10}_{-14.17}$ & $[0.0,\,45.8]$ \\
51 Eri & b & $55.60^{+23.40}_{-10.40}$ & $33.00^{+3.19}_{-7.96}$ & $25.70^{+24.62}_{-10.60}$ & $[3.4,\,61.9]$ \\
AB Pic & b & $55.70^{+15.70}_{-8.90}$ & $88.70^{+0.95}_{-0.80}$ & $31.80^{+10.47}_{-15.50}$ & $[0.0,\,44.2]$ \\
AF Lep & b & $61.50^{+15.70}_{-8.70}$ & $57.80^{+0.65}_{-0.60}$ & $0.00^{+15.51}_{-0.00}$ & $[0.0,\,29.5]$ \\
BD-210397 & c & $66.50^{+22.00}_{-9.40}$ & $34.00^{+4.16}_{-3.69}$ & $31.90^{+19.15}_{-14.08}$ & $[2.9,\,57.6]$ \\
beta Pic & c & $88.80^{+0.12}_{-0.12}$ & $88.60^{+0.79}_{-0.67}$ & $0.00^{+0.64}_{-0.00}$ & $[0.0,\,1.2]$ \\
beta Pic & b & $88.80^{+0.12}_{-0.12}$ & $89.00^{+0.14}_{-0.14}$ & $0.00^{+0.31}_{-0.00}$ & $[0.0,\,0.5]$ \\
beta Pic & d & $88.80^{+0.12}_{-0.12}$ & $88.80^{+0.99}_{-0.71}$ & $0.00^{+0.78}_{-0.00}$ & $[0.0,\,1.6]$ \\
eps Eri & b & $52.10^{+19.50}_{-9.90}$ & $39.00^{+6.75}_{-5.42}$ & $10.50^{+17.03}_{-10.50}$ & $[0.0,\,47.8]$ \\
GJ 504 & b & $18.90^{+2.60}_{-2.10}$ & $43.40^{+11.97}_{-12.73}$ & $23.80^{+11.67}_{-11.01}$ & $[0.0,\,36.8]$ \\
\hline
\end{longtable*}
\endgroup

\clearpage


\onecolumngrid
\clearpage
\section{Summary of relative astrometries and radial velocities used in orbital fitting}

\begin{table}[h]
\centering
\begin{threeparttable}
\caption{\textbf{Summary of imaging relative astrometry used in orbital fitting}. Only the first 10 rows are shown; the full table will be available upon publication.}
\label{tab:rel_astr}

\footnotesize
\setlength{\tabcolsep}{3pt}
\begin{tabular}{@{}llll@{}}
\hline
\hline
Name & \shortstack[l]{Number of\\Relative\\Astrometry} & Observation Baseline & Reference \\
\hline
1RXS J034231.8+121622 B & 9 & 2007-12-14 to 2023-02-08 & 1,2,3,4 \\
2MASS J01225093$-$2439505 B & 9 & 2012-10-12 to 2019-06-17 & 3,5,6,7 \\
2MASS J23513366+3127229 B & 11 & 2011-06-21 to 2019-07-07 & 3,4,8,9 \\
51 Eri b & 21 & 2014-12-18 to 2023-10-18 & 10,11,12 \\
AB Pic b & 7 & 2003-03-17 to 2023-10-26 & 13,14,15,16 \\
AF Lep b & 13 & 2011-10-21 to 2024-01-14 & 17,18,19,20,21,22 \\
$\beta$ Pic b & 40 & 2003-11-10 to 2018-11-19 & 23,24,25,26,27 \\
$\beta$ Pic c & 3 & 2020-02-10 to 2020-03-08 & 28 \\
$\beta$ Pic d & 8 & 2014-12-08 to 2026-03-29 & 29,30 \\
GJ 504 b & 17 & 2011-03-26 to 2023-07-04 & 31,32,33 \\
\hline
\end{tabular}

\begin{tablenotes}[flushleft]
\footnotesize
\item \textit{Note.}
1. \cite{Janson2012},
2. \cite{Janson2014},
3. \cite{Bowler2015a},
4. \cite{Bowler2020},
5. \cite{Bryan2016},
6. \cite{Bryan2020},
7. \cite{Bowler2013},
8. \cite{Bowler2012},
9. \cite{Sanghi2024},
10. \cite{DeRosa2015},
11. \cite{Maire2019},
12. \cite{Balmer2025},
13. \cite{Chauvin2005},
14. \cite{Rameau2013},
15. \cite{PalmaBifani2023},
16. \cite{Gandhi2025},
17. \cite{Mesa2023},
18. \cite{DeRosa2023},
19. \cite{Franson2023},
20. \cite{Balmer2025AFLEP},
21. \cite{Bonse2025},
22. \cite{Franson2024},
23. \cite{Lagrange2009},
24. \cite{Chauvin2012},
25. \cite{Nielsen2014},
26. \cite{Wang2016},
27. \cite{Lagrange2019},
28. \cite{Nowak2020},
29. \cite{Sutlieff2026},
30. \cite{Gibbs2026},
31. \cite{Kuzuhara2013},
32. \cite{Bonnefoy2018},
33. \cite{Malin2025},
34. \cite{Janson2006},
35. \cite{Neuhauser2005},
36. \cite{Neuhauser2008},
37. \cite{Ginski2014},
38. \cite{Kraus2014},
39. \cite{Stolker2021},
40. \cite{Konopacky2016},
41. \cite{Maire2018},
42. \cite{Roberts2025},
43. \cite{Biller2010},
44. \cite{Mugrauer2010},
45. \cite{Mugrauer2012},
46. \cite{Biller2013},
47. \cite{Maire2016},
48. \cite{Beust2016},
49. \cite{MussoBarcucci2019},
50. \cite{FransonBowler2023},
51. \cite{Ireland2011},
52. \cite{Pearce2019},
53. \cite{Metchev2004},
54. \cite{Serabyn2009},
55. \cite{Bottom2017},
56. \cite{Carson2013},
57. \cite{Currie2018},
58. \cite{Uyama2020},
59. \cite{Chauvin2017},
60. \cite{Cheetham2019},
61. \cite{Stolker2020},
62. \cite{Petrus2021},
63. \cite{Blunt2023},
64. \cite{Carter2023},
65. \cite{Ginski2013},
66. \cite{Milli2017},
67. \cite{Delorme2017},
68. \cite{WardDuong2021},
69. \cite{Kral2026},
70. \cite{Hinkley2023},
71. \cite{Currie2023},
72. \cite{Winterhalder2025},
73. \cite{Bovie2025},
74. \cite{Balmer2026},
75. \cite{Wahhaj2011},
76. \cite{Wang2026},
77. \cite{Ratzka2005},
78. \cite{Hayoz2026}.
\end{tablenotes}
\end{threeparttable}
\end{table}

\clearpage

\begin{table}[h]
\centering
\begin{threeparttable}
\caption{\textbf{Summary of radial-velocity data used in orbital fitting}. Only the first 10 rows are shown; the full table will be available upon publication.}
\label{tab:rv_data}

\footnotesize
\setlength{\tabcolsep}{9pt}
\begin{tabular}{@{}llll@{}}
\hline
\hline
Name & \shortstack[l]{Number of\\RVs} & Observation Baseline & Reference \\
\hline
$\beta$ Pic & 90 & 2010-12-12 to 2021-01-29 & 2 \\
$\epsilon$ Eri & 509 & 1987-09-08 to 2020-12-06 & 3,4,5 \\
GJ 832 & 274 & 1998-08-08 to 2020-01-01 & 1,6,7 \\
HD 10647 & 338 & 1998-01-17 to 2020-01-17 & 1,8,9 \\
HD 108202 & 43 & 2005-05-21 to 2022-03-15 & 1,10 \\
HD 109286 & 45 & 2012-04-22 to 2018-04-10 & 11 \\
HD 111998 & 149 & 2006-03-12 to 2014-05-21 & 12 \\
HD 113337 & 207 & 2007-01-23 to 2022-07-15 & 13 \\
HD 128717 & 135 & 2022-06-05 to 2026-03-16 & 14 \\
HD 137510 & 52 & 2001-03-30 to 2008-06-06 & 3 \\
\hline
\end{tabular}

\begin{tablenotes}[flushleft]
\footnotesize
\item \textit{Note.}
1. \cite{Perdelwitz2024};
2. \cite{Lagrange2019RV};
3. \cite{Fischer2014};
4. \cite{Mawet2019};
5. \cite{LlopSayson2021};
6. \cite{Bailey2009};
7. \cite{Wittenmyer2014};
8. \cite{Butler2006};
9. \cite{Marmier2013};
10. \cite{Philipot2023RV};
11. \cite{Demangeon2021};
12. \cite{Borgniet2017};
13. \cite{Borgniet2014};
14. \cite{Pinamonti2026};
15. \cite{TalOr2019};
16. \cite{Mayor2004};
17. \cite{Boisse2012};
18. \cite{Feng2022};
19. \cite{Segransan2010};
20. \cite{Butler2017};
21. \cite{Jones2002};
22. \cite{Santos2001};
23. \cite{LoCurto2010};
24. \cite{Jenkins2017};
25. \cite{Kiefer2019};
26. \cite{Rickman2019};
27. \cite{Diaz2012};
28. \cite{DeRosa2020PiMen};
29. \cite{Rosenthal2021};
30. \cite{Udry2003};
31. \cite{Zhang2025};
32. \cite{Rodigas2016};
33. \cite{Sozzetti2006};
34. \cite{Fischer2009};
35. \cite{Segransan2011};
36. \cite{Franson2022RV};
37. \cite{Pena2026};
38. \cite{Barbieri2023}.
\end{tablenotes}
\end{threeparttable}
\end{table}

\clearpage


\clearpage

\bibliography{sample701}{}
\bibliographystyle{aasjournalv7}



\end{document}